\documentclass[twocolumn,twocolappendix]{aastex63}

\received{June 1, 2019}
\revised{January 10, 2019}
\accepted{\today}
\submitjournal{ApJ}

\shortauthors{Su\'arez et al.}
\shorttitle{HN Peg B: the L/T Transition at 300 Myr}
\graphicspath{{./}{figures/}}

\makeatletter
\newcommand\notsotiny{\@setfontsize\notsotiny\@vipt\@viipt}
\makeatother

\usepackage[caption=false]{subfig} 
\usepackage[referable]{threeparttablex} 
\usepackage{booktabs} 
\usepackage{multirow} 
\usepackage{float}
\usepackage{amsmath}
\usepackage{chngpage} 
\usepackage{ragged2e}
\usepackage{tabularx}

\begin{document}

\title{Ultracool Dwarfs Observed with the Spitzer Infrared Spectrograph. I. An Accurate Look at the L-to-T Transition at $\sim$300 Myr from Optical through Mid-infrared Spectrophotometry}

\correspondingauthor{Genaro Su\'arez}
\email{gsuarez@uwo.ca}

\author[0000-0002-2011-4924]{Genaro Su\'arez}
\affil{Department of Physics and Astronomy, Institute for Earth and Space Exploration, The University of Western Ontario, 1151 Richmond St, London, ON N6G 1N9, Canada}

\author[0000-0003-3050-8203]{Stanimir Metchev}
\affil{Department of Physics and Astronomy, Institute for Earth and Space Exploration, The University of Western Ontario, 1151 Richmond St, London, ON N6G 1N9, Canada}
\affil{Department of Astrophysics, American Museum of Natural History, 200 Central Park West, New York, NY 10024--5102} 

\author[0000-0002-3681-2989]{Sandy K. Leggett}
\affil{Gemini Observatory, Northern Operations Center, 670 N. A’ohoku Place, Hilo, HI 96720}

\author[0000-0001-6800-3505]{Didier Saumon}
\affil{Los Alamos National Laboratory, P.O. Box 1663, Los Alamos, NM 87545}

\author[0000-0002-5251-2943]{Mark S. Marley}
\affil{NASA Ames Research Center, Mail Stop 245-3, Moffett Field, CA 94035, USA}



\begin{abstract}
We present $Spitzer$ IRS 5--14~$\mu$m spectra and 16~$\micron$ and 22~$\micron$ photometry of the 
T2.5 companion to the $\sim$300 Myr-old G0V star HN~Peg. We incorporate previous 0.8--5~$\micron$ observations to obtain the most comprehensive spectral energy distribution of an intermediate-gravity L/T-transition dwarf which, together with an accurate Gaia EDR3 parallax of the primary, enable us to derive precise fundamental parameters. We find that young ($\approx$0.1--0.3~Gyr) early-T dwarfs on average have $\approx$140~K lower effective temperatures, $\approx$20\% larger radii, and similar bolometric luminosities compared to $\gtrsim$1~Gyr-old field dwarfs with similar spectral types. Our accurate infrared spectrophotometry offers new detail at wavelengths where the dominant carbon-bearing molecules have their strongest transitions: at 3.4~$\micron$ for methane and at 4.6~$\micron$ for carbon monoxide. We assess the performance of various widely available photospheric models and find that models with condensates and/or clouds better reproduce the full SED of this moderately young early-T dwarf. However, cloud-free models incorporating a more general convective instability treatment reproduce at least the low-resolution near-IR spectrum similarly well. Our analysis of $R\approx2300$ $J$-band spectra shows that the near-infrared potassium absorption lines in HN~Peg~B have similar strengths to those seen in both younger and older T2-T3 dwarfs. We conclude that while alkali lines are well-established as surface gravity indicators for L-type or warmer stars, they are insensitive to surface gravity in early-T dwarfs.
 
\end{abstract}


\section{INTRODUCTION}
The transition from the L to T spectral type is marked by the settling of dust condensates and the replacement of carbon monoxide with methane as the dominant carbon-bearing absorber in substellar atmospheres.  While the processes are seemingly concurrent, studies of young L-to-T transition dwarfs have revealed that their onset depends on the age, and hence, the surface gravity of the dwarf. Like other young brown dwarfs, young L/T-transition dwarfs are redder than their $>$1~Gyr-old field counterparts \citep[e.g.;][]{Kirkpatrick_etal2008}.  The discovery of the first young L/T-transition dwarfs---HD~203030~B  \citep[L7.5;][]{Metchev-Hillenbrand2006} and HN~Peg~B \citep[T2.5;][]{Luhman_etal2007}---also showed that they are cooler compared to their older counterparts.  Their effective temperatures are more consistent with those of field T6 dwarfs.  Hence, both the dust sedimentation and the appearance of methane in the spectra are delayed at young ages or, equivalently \citep[e.g.,][]{Burrows_etal1997,Chabrier_etal2000}, at low surface gravities. \citet{Marley_etal2012} explain this as a result of the higher opacity of lower-gravity atmospheres, since the condensation of volatiles occurs at lower pressures (higher up in the atmospheres), so clouds remain visible to cooler temperatures for lower-gravity objects.

The atmospheres of young late-L and T dwarfs are relevant to understanding exoplanetary atmospheres.  Most of the directly imaged extrasolar giant planets have the appearance of young L or T dwarfs: e.g., HR~8799~bcde \citep{Marois_etal2008,Marois_etal2010}, HD~95086~b \citep{Rameau_etal2013, Rameau_etal2013b}, 51~Eri~b \citep{Macintosh_etal2015}, and YSES 2b \citep{Bohn_etal2021}.  Both HD~203030~B and HN~Peg~B themselves likely have planet-like masses \citep{Metchev-Hillenbrand2006,Luhman_etal2007,Leggett_etal2008,Miles-Paez_etal2017}.  Of the two, HN~Peg~B is much more widely separated ($43\arcsec$) from its host star, allowing uncontaminated spectroscopy, and is the only young ($\approx$300 Myr) substellar companion for which spectra with the $Spitzer$ Infrared Spectrograph (IRS) were obtained during the $Spitzer$ Cryogenic Mission. The heretofore adopted age for HN~Peg~B assumes co-evality with the primary, for which \citet{Luhman_2007} estimated an age of $300\pm200$~Myr based on its lithium abundance, rotation, chromospheric activity, and kinematics.

Herein we assemble previously published ground-based (IRTF/SpeX, Keck/NIRSPEC, and Gemini/NIRI) with new unpublished $Spitzer$ IRS observations to present nearly complete spectroscopic coverage of HN~Peg~B between 0.8--15~$\micron$ (Section \ref{sec:data}).  We complement this with additional optical through mid-infrared photometry from Pan-STARRS, the IRTF, and $Spitzer$ IRAC and IRS out to 22~$\micron$.  We compare the data to other T2--T3 dwarfs and to suites of widely-used atmospheric models, and use the spectrophotometry to refine the fundamental parameters of the brown dwarf and to obtain a stand-alone age estimate (Section~\ref{sec:results}). We discuss our principal findings and conclude in Sections~\ref{sec:discussion} and \ref{sec:conclusions}.

\section{OBSERVATIONS AND DATA REDUCTION}
\label{sec:data}
All photometric and spectroscopic observations are summarized in Table~\ref{tab:observations}.  The corresponding photometric measurements are listed in Table \ref{tab:HN_PegB_phot} and the spectra are available in the online journal.  Below we describe the published and newly-reported observations.

\begin{table*}
\caption{Summary of Observations.}
  \scriptsize
  \centering
  \label{tab:observations}
  \begin{threeparttable}
 	\begin{tabular}{@{\extracolsep{2pt}}ccccc@{}}
    \toprule
	Telescope/Instrument & Band$^1$/Wavelength  & Resolution & Observation               & Reference \\
	                     &                      &            & (UT Date)               &           \\
    \midrule                                                                      
	\multicolumn{4}{c}{\textbf{Photometry}}                                             \\
    \midrule                                                                      
	PS1/GPC1             & $z_{P1}$ and $y_{P1}$                             & \nodata & 2010--2015                & 1         \\
	IRTF/SpeX            & $J_{\rm{MKO}}$, $H_{\rm{MKO}}$ and $K_{\rm{MKO}}$ & \nodata & 2006 Jul 2                & 2         \\
	{\it Spitzer}/IRAC   & Channels 1--4                                     & \nodata & 2004 Jun 10               & 2         \\
	{\it Spitzer}/IRS    & Blue and Red Peak-up                              & \nodata & 2007 Dec 8                & 3         \\
	\\
    \midrule                                                                      
	\multicolumn{4}{c}{\textbf{Spectroscopy}}                                           \\
    \midrule                                                                      
	IRTF/SpeX            & 0.8--2.5 $\micron$    & $\approx$100  & 2006 Jun 15               & 2         \\
	Keck/NIRSPEC         & 1.15--1.38 $\micron$  & $\approx$2300 & 2008 Jul 8                & 4         \\
    Gemini/NIRI          & 2.96--4.07 $\micron$  & $\approx$400  & 2007 Jul 26, 30; Aug 1, 5 & 5         \\
	{\it Spitzer}/IRS    & 5.2--14.2 $\micron$   & $\sim$60--130 & 2008 Jul 11; Aug 2        & 3         \\
    \bottomrule
	\end{tabular}
	\begin{tablenotes}[para,flushleft]
	{\bf Reference:} (1) \citet{Chambers_etal2016}; (2) \citet{Luhman_etal2007}; (3) this paper; (4) \citet{Zhou_etal2018}; (5) \citet{Leggett_etal2008}.\\
	$^1$The effective wavelengths and band widths of the filters are listed in Table \ref{tab:HN_PegB_phot}.
	\end{tablenotes}
 \end{threeparttable}
\end{table*}

\begin{table}
\caption{New and Compiled Photometry of HN~Peg~B.}
  \scriptsize
  \centering
  \label{tab:HN_PegB_phot}
  \begin{threeparttable}
	\begin{tabularx}{\linewidth}{@{\extracolsep{-4pt}}cccccc@{}}
    \toprule
	Telescope/Instrument                & Band           & $\lambda_{\rm{eff}}^1$  & W$_{eff}^1$ & Magnitude & Flux      \\
	                                    &                & ($\mu$m)                & ($\mu$m)  & (mag)     & ($\mu$Jy) \\
	\midrule                                                                      
    PS1/GPC1                            & $z_{P1}$       & 0.890	               & 0.097     & $19.65\pm0.12$ & $31\pm3$   \\
	                                    & $y_{P1}$       & 0.976	               & 0.061     & $18.10\pm0.08$ & $127\pm9$  \\
	IRTF/SpeX                           & $J_{\rm{MKO}}$ & 1.257	               & 0.147     & $15.86\pm0.03$ & $707\pm20$ \\
	                                    & $H_{\rm{MKO}}$ & 1.632	               & 0.276     & $15.40\pm0.03$ & $720\pm20$ \\
	                                    & $K_{\rm{MKO}}$ & 2.165	               & 0.326     & $15.12\pm0.03$ & $578\pm16$ \\
	{\it Spitzer}/IRAC                  & $CH1$          & 3.679	               & 0.684     & $13.72\pm0.04$ & $913\pm36$ \\
	                                    & $CH2$          & 4.309	               & 0.865     & $13.39\pm0.02$ & $792\pm19$ \\
	                                    & $CH3$          & 5.677	               & 1.256     & $13.08\pm0.10$ & $674\pm63$ \\
	                                    & $CH4$          & 7.864	               & 2.529     & $12.58\pm0.11$ & $603\pm62$ \\
	{\it Spitzer}/IRS                   & Blue           & 15.30                   & 4.677     & $11.78\pm0.07$ & $305\pm21$ \\
	                                    & Red            & 21.58                   & 6.998     & $11.83\pm0.18$ & $143\pm24$ \\
	\bottomrule
	\end{tabularx}
	\begin{tablenotes}[para,flushleft]
        $^1$The filter effective wavelengths and widths were obtained by convolving the filter response curves with the best-fitting BT-Settl atmosphere to the HN~Peg~B SED (Figure~\ref{fig:SED}).
	\end{tablenotes}
 \end{threeparttable}
\end{table}

\subsection{Published Photometry and Spectroscopy}
The discovery paper of HN~Peg~B \citep{Luhman_etal2007} included Mauna Kea Observatory (MKO) $J$-, $H$-, and $K$-band photometry from the imaging camera of the NASA IRTF/SpeX spectrograph \citep{Rayner_etal2003} and 3.6--8.0~$\micron$ four-band photometry from {\it Spitzer}/IRAC \citep{Fazio_etal2004}.  Additional red-optical photometric measurements in the $z_{P1}$ and $y_{P1}$ filters are available from Pan-STARRS release 1 \citep[PS1;][]{Chambers_etal2016}. Photometry of HN~Peg~B is also available from 2MASS \citep{Skrutskie_etal2006} and \textit{WISE} \citep{Cutri_etal2013}. However, the 2MASS and \textit{WISE} fluxes are potentially contaminated by the bright primary star, so we do not consider them here.

Spectra of HN~Peg~B have been published in the discovery paper by \citet[][$R\approx100$, 0.8--2.5~$\micron$ with SpeX on the IRTF]{Luhman_etal2007} and \citet[][$R\approx400$, 3--4~$\micron$ with NIRI on Gemini North]{Leggett_etal2008}. We further obtained an $R\approx2300$ 1.143--1.375~$\micron$ spectrum with NIRSPEC on Keck II, that we published in collaboration with \citet{Zhou_etal2018}.  We present an updated extraction and analysis of this spectrum below.

\subsubsection{Keck/NIRSPEC $J$-band Spectrum}
\label{sec:NIRSPEC_HN_PegB}
A total of eight individual 300 s exposures of HN~Peg~B were obtained following a standard ABBA nodding observing strategy along the slit. We used the two-pixel ($0\farcs38$) wide slit and the N3 (1.143--1.375 $\mu$m) filter; hat is similar to a near-IR $J$-band filter. The A0 standard star HD~216308 was observed at a similar airmass.  Ne and Ar arc lamp spectra and flat-field lamp spectra were obtained after the science target.

For the data reduction, we first removed cosmic rays of the individual target frames using the L.A. Cosmic algorithm \citep{vanDokkum2001}. We then used the REDSPEC pipeline \citep{McLean_etal2003} for the following steps: $i)$ to spatially rectify and wavelength-calibrate the spectra, $ii)$ to subtract pairs of nodded frames from each other to remove the sky background and to divide them by the flat field, and $iii)$ to extract both the target spectrum from each nod position and a background spectrum along nearby apertures parallel to the nod traces in the pairwise-subtracted frames. We subtracted the background spectra from the target spectra to remove residual artifacts from incomplete sky subtraction in the pairwise nodding or saturated OH emission lines.

The telluric A0 standard star spectra were extracted using the same prescription, divided by a 9500~K blackbody  \citep{Schmidt-Kaler1982}, and normalized by the median of the resulting spectrum. The target spectra were divided by the normalized standard-star spectra to correct for telluric absorption and instrument transmission features.

\begin{figure}
	\includegraphics[width=0.5\textwidth]{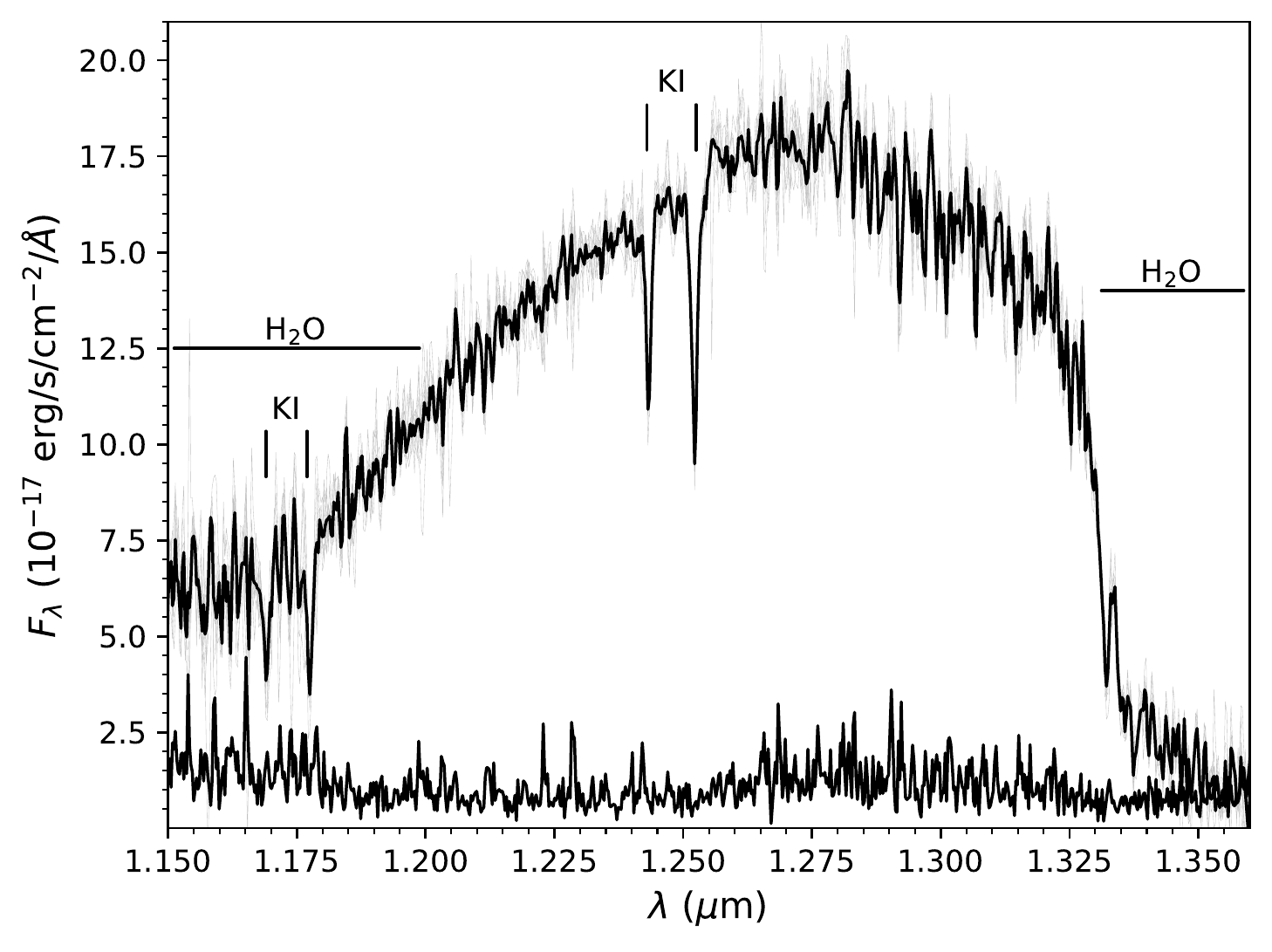}
	\caption{$J$-band (N3 filter) $R\approx2300$ NIRSPEC spectrum of HN~Peg~B.  The median-combined spectrum is plotted in black and the individual spectra in gray.  Flux uncertainties are shown along the bottom. The main spectral features are indicated.}
	\label{fig:NIRSPEC_spectrum}
\end{figure}

The wavelength-calibrated and telluric-corrected individual spectra were median-combined using the IRAF {\sc scombine} task. The uncertainties of the combined spectrum correspond to the 68\% central confidence interval. We used the MKO $J$-band photometry from \citet{Luhman_etal2007} to flux-calibrate the median-combined spectrum. The final NIRSPEC $J$-band spectrum of HN~Peg~B is shown in Figure \ref{fig:NIRSPEC_spectrum}.

\subsection{New Spitzer IRS Photometry and Spectroscopy}
We observed HN~Peg~B with the Infrared Spectrograph \citep[IRS;][]{Houck_etal2004} on the {\it Spitzer Space Telescope} as part of {\it Spitzer} GO program 40489 (PI: Metchev). Photometric measurements were obtained in the blue and red peak-up imaging channels in IRS.  Spectroscopy was obtained in the 5.2--14.2~$\micron$ short-low resolution IRS mode.

\subsubsection{IRS 13--26~$\micron$ Peak-Up Photometry}
\begin{figure*}
\includegraphics[width=1.0\textwidth]{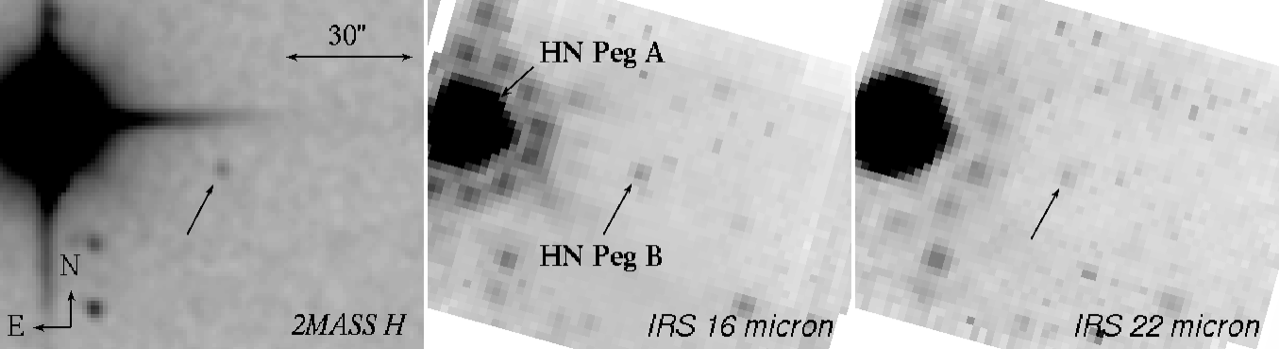}
\caption{Images of the HN~Peg~B substellar companion from 2MASS (left) and from the {\it Spitzer} IRS blue (middle) and red (right) peak-up arrays. }
\label{fig:IRS_images}
\end{figure*}

The blue (13.3--18.7~$\micron$) and red (18.5--26.0~$\micron$) peak-up array images of HN~Peg~B were obtained in nine sets of twelve 30~s exposures, with each set scattered along a large ($\sim$160$\arcsec$-wide) random pattern with half-integer pixel offsets. The total exposure time was 54~min in each of the two peak-up arrays. The post-BCD images of HN~Peg~B in the blue and red peak-up arrays are shown in Figure~\ref{fig:IRS_images}.  

We performed photometry with custom IDL routines on the BCD images. To improve the SNR of HN~Peg~B in each flux measurement, we divided the 108 individual blue peak-up BCD frames into twelve sets of nine frames, and each set was median-combined to produce a single image. The 108 red peak-up BCD frames were similarly grouped and median-combined into four independent images, each the result of 27 frames. Our fiducial flux estimate in each filter is the mean of the independent measurements. We used an $r=3$~pix and an $r=4$~pix aperture for photometry in the blue and read peak-up images, respectively. (The \textit{Spitzer} IRS peak-up imager pixel scale is $\approx$1$\farcs$84 pix$^{-1}$.) Since the $>$11~$\micron$ spectrum of HN~Peg~B is approximately blackbody-like (see Figure \ref{fig:IRS_spectrum}), we used aperture corrections of 1.562 and 1.575 for the blue and red peak-up photometry, as specified for the respective aperture radii and 2000~K blackbodies in the IRS Instrument Handbook (Table 4.11\footnote{\url{http://irsa.ipac.caltech.edu/data/SPITZER/docs/irs/}}). We estimate the photometric uncertainty in each filter by combining in quadrature the scatter (few percent) of the individual flux measurements and a $\sim$10\% empirically determined uncertainty in the local background estimation, based on experiments with annuli of various radii (between 4 and 7 pixels) and widths (from 1 to 3 pixels).

\subsubsection{IRS Low-Resolution 5.2--14.2~$\micron$ Spectra}
We used the SL2 (5.2--8.7~$\micron$) and SL1 (7.4--14.5~$\micron$) modes in IRS to obtain $R\sim$ 60--130 spectra of HN~Peg~B. We dithered along two positions of the 3$\farcs$7-wide slit for subsequent background removal through pairwise subtraction. Exposures were 60~s per pointing, with a total of 128 pairs in SL2 and 224 pairs in SL1 over two separate AOR observations on 11 July and 2 August 2008. The total exposure times were 4.27 hours in SL2 and 7.47 hours in SL1.

\begin{figure}
	\includegraphics[width=0.5\textwidth]{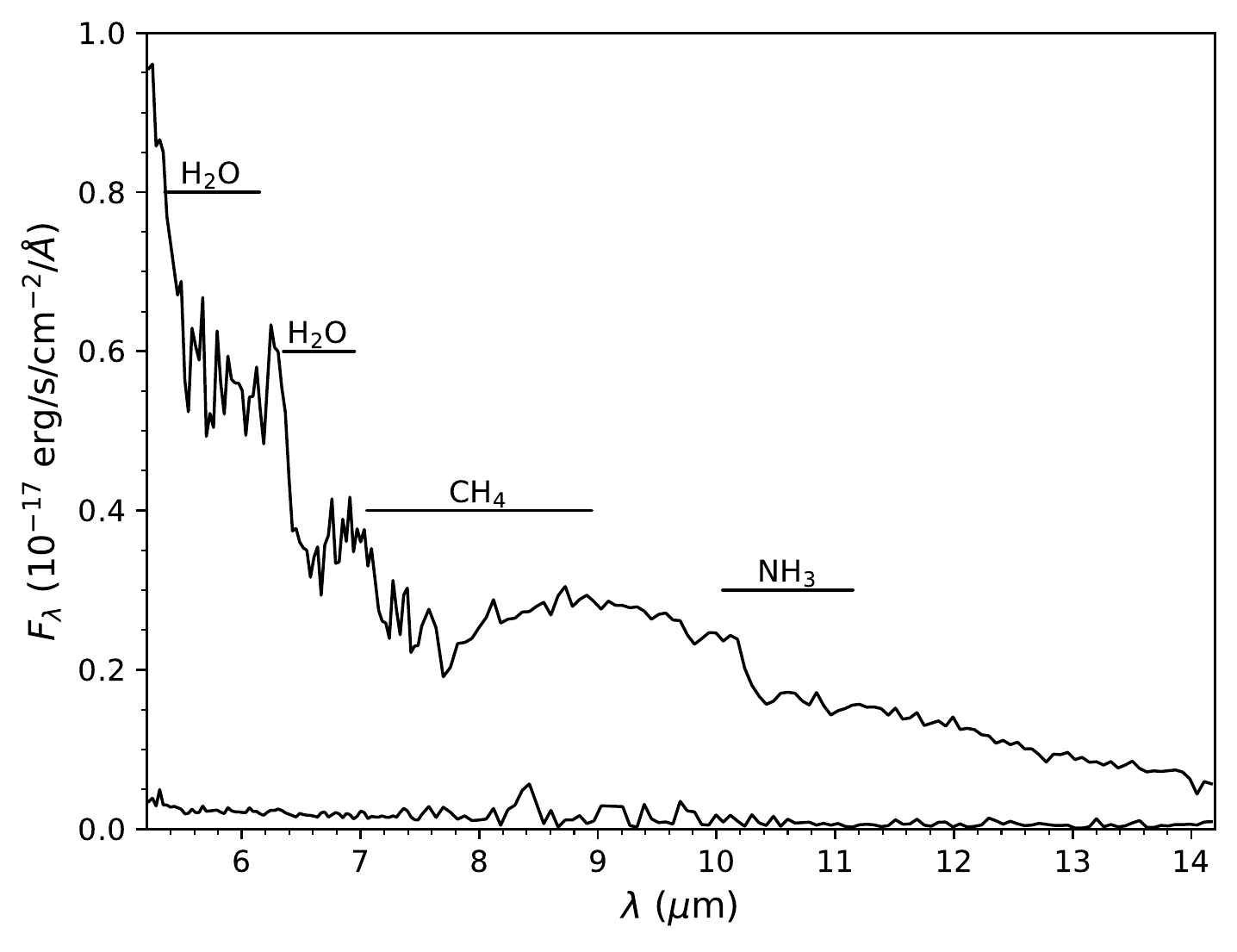}
	\caption{$Spitzer$ IRS $R\sim60$--130 spectrum of HN~Peg~B together with flux uncertainties along the bottom. The main molecular absorption bands present in the spectrum are indicated.}
	\label{fig:IRS_spectrum}
\end{figure}

All IRS data were reduced with the S18.7.0 version of the {\it Spitzer} Science Center (SSC) data reduction pipeline, which produces sets of basic calibrated data (BCD) files for each exposure. We used IRSCLEAN (v1.9) with the appropriate IRS campaign masks to clean rogue pixels, iterating once with additional rogue pixels marked by hand for more aggressive outlier rejection. We then median-combined the images taken at each nod, and pairwise subtracted them. We used SPICE (v2.2) to optimally extract and average the two spectral traces. We flux-calibrated the spectrum with respect to the measured IRAC channel 4 flux from \citet{Luhman_etal2007}, following the approach described in \citet{Cushing_etal2006}. The reduced 5.2--14.2~$\micron$ spectrum of HN~Peg~B is shown in Figure~\ref{fig:IRS_spectrum}.

\section{RESULTS AND ANALYSIS}
\label{sec:results}
We first compare our $R\approx2300$ NIRSPEC $J$-band spectrum of HN~Peg~B (T2.5) to NIRSPEC spectra of other young or old field T2--T3 dwarfs to examine the surface gravity dependence of the potassium absorption lines (Section~\ref{sec:results:ki}). Then, we compare the assembled spectroscopy and photometry to widely-used photospheric models to determine the effective temperature and surface gravity of HN~Peg~B, and to assess various models' performance (Section~\ref{sec:model_fits}). Finally, we use the comprehensive spectral energy distribution (SED) of HN~Peg~B and the accurately known \textit{Gaia} parallax of HN~Peg~A to derive precise fundamental parameters (Section~\ref{sec:phy_par}).

\subsection{The $J$-band \ion{K}{1} Doublet in Early-T Dwarfs: No Sensitivity to Surface Gravity}
\label{sec:results:ki}

The strengths of the \ion{K}{1} doublets at 1.17~$\micron$ and 1.25~$\mu$m are known to decrease with decreasing surface gravity late-M and early-L dwarfs \citep{McGovern_etal2004}.  The behavior continues into the late-L dwarfs, as observed by \citet{Miles-Paez_etal2017} in $R\approx2300$ $J$-band spectra of young and old L7--L8 dwarfs.  Hence, lower surface gravity leads to weaker \ion{K}{1} lines in all M and L dwarfs.  However, \citet{Knapp_etal2004} discuss the opposite effect in late-T dwarfs.  They find increasing near-infrared \ion{K}{1} absorption strengths with decreasing surface gravity in both observed and model spectra of T6 and T8 dwarfs.

It is therefore likely that over some intermediate spectral type range, potentially between early- and mid-T dwarfs, the potassium line strengths---an otherwise excellent surface gravity proxy---are not usable as such.  Based on their low-resolution ($R\sim100$) SpeX spectrum of HN~Peg~B, \citet{Luhman_etal2007} report somewhat weaker \ion{K}{1} lines strengths.  Given that HN~Peg~B is moderately young, this could indicate a continuation of the positive correlation between potassium line strengths and surface gravity from L dwarfs into at least early-T dwarfs.  However, as we demonstrate below, we do not confirm this behavior in our higher resolution ($R\approx2300$) NIRSPEC spectrum of HN~Peg~B.

To investigate the effect of surface gravity on the potassium line strengths in early-T dwarfs, we compare our NIRSPEC $J$-band (N3 filter) spectrum of HN~Peg~B to archival NIRSPEC spectra of four other T2--T3 dwarfs with younger or older ages. The comparison objects are SDSS J125453.90$-$012247.4 \citep[SDSS 1254$-$01;][]{Leggett_etal2000}, 2MASS J11061197+2754225 \citep[2MASS 1106+27;][]{Looper_etal2007}, SIMP J013656.6+093347 \citep[SIMP 0136+09;][]{Artigau_etal2006}, and SDSS J102109.69$-$030420.1 \citep[SDSS 1021$-$03;][]{Leggett_etal2000}. Table \ref{tab:dwarfs_comparison} lists the spectral types, ages, surface gravities, and the respective references for these objects. The surface gravities are either estimates from the respective studies \citep[][this work]{Filippazzo_etal2015,Gagne_etal2017}, or as expected from evolutionary models \citep{Saumon-Marley2008,Marley_etal2018,Phillips_etal2020} based on the assumed ages.

\begin{figure*}
	\centering
	\subfloat{\includegraphics[width=0.5\textwidth]{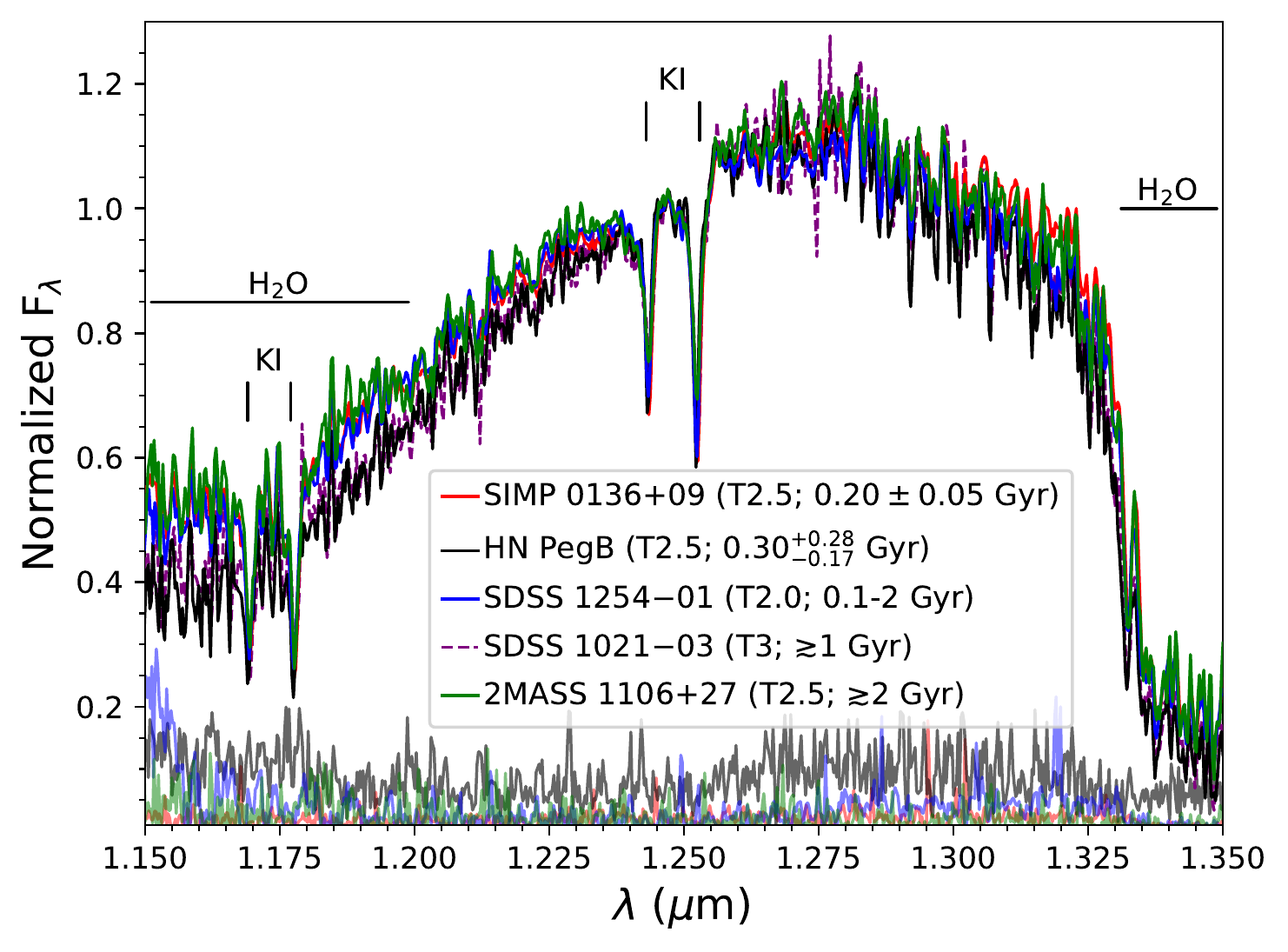}}
	\subfloat{\includegraphics[width=0.5\textwidth]{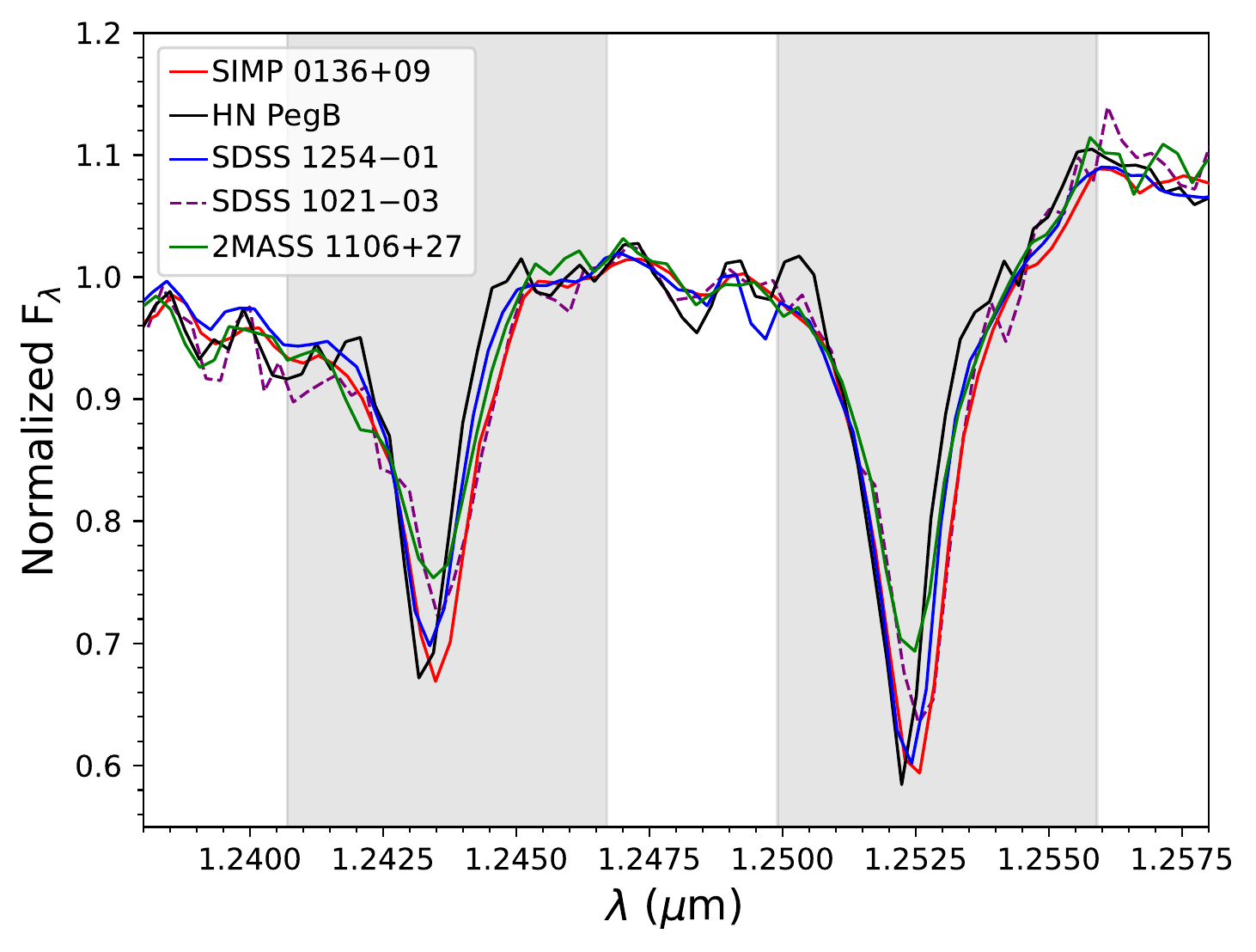}} \\
	\subfloat{\includegraphics[width=0.5\textwidth]{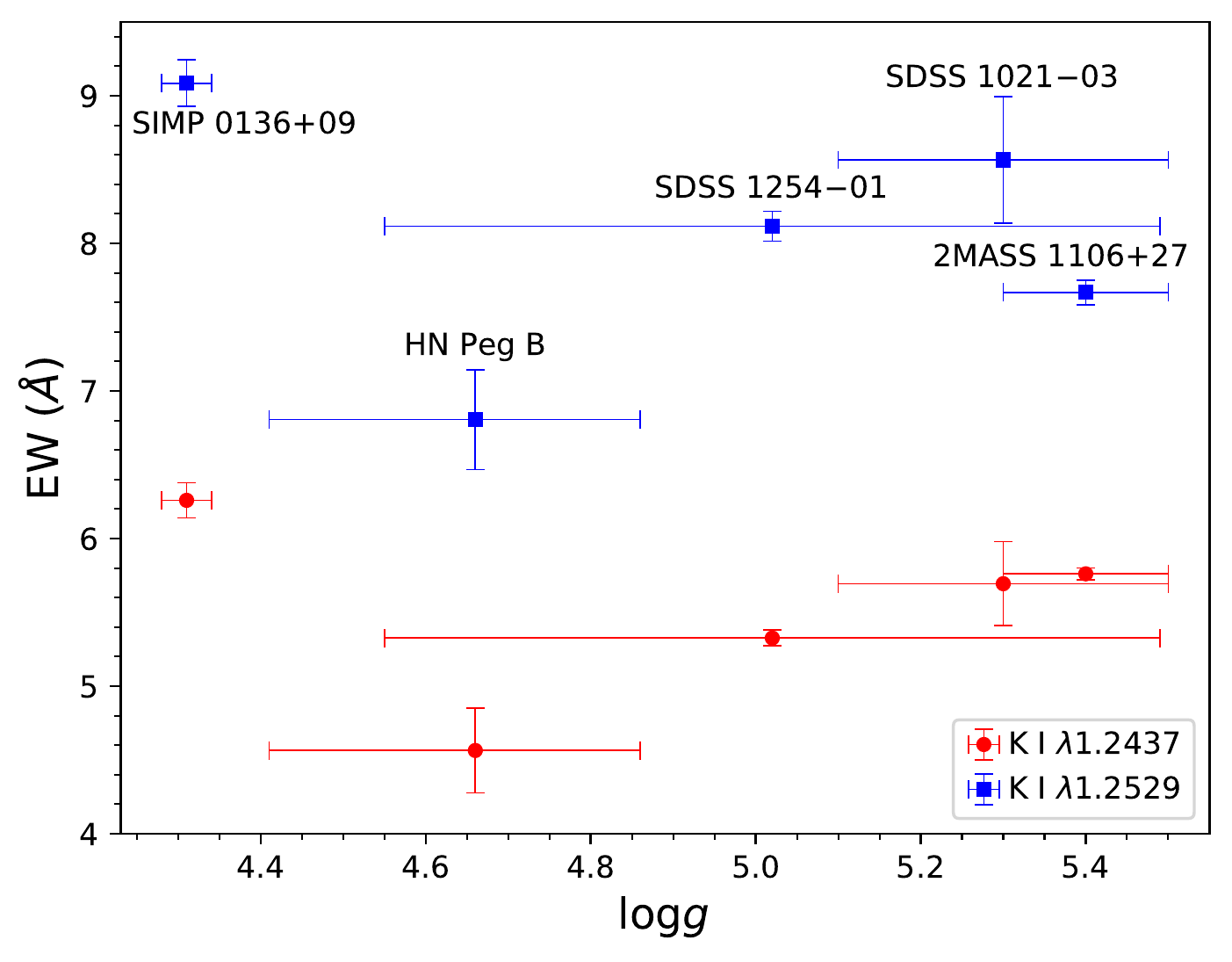}}
	\caption{{\bf Top left panel:} Comparison of the NIRSPEC N3-band 1.15--1.35~$\micron$ spectrum of HN~Peg~B (T2.5) to spectra of younger and older T2--T3 dwarfs. The spectra are normalized to an average of unity in the 1.246--1.250 $\mu$m wavelength region, between the two \ion{K}{1} absorption lines. {\bf Top right panel}: A zoom on to the \ion{K}{1} doublet lines at 1.2437~$\mu$m and 1.2529~$\mu$m.  The shaded regions correspond to the wavelength ranges over which we calculated the EWs of the \ion{K}{1} lines. The targets are listed in the legend in order of increasing surface gravity (see Table~\ref{tab:dwarfs_comparison}). {\bf Bottom panel:} EWs of the $J$-band \ion{K}{1} doublet as a function of surface gravity for HN~Peg~B and the four comparison T2--T3 dwarfs.}
	\label{fig:dwarfs_comparison}
\end{figure*}

We retrieved all observations of the comparison T2--T3 dwarfs, along with observations of the corresponding telluric standards, from the Keck Observatory Archive\footnote{\url{https://koa.ipac.caltech.edu/cgi-bin/KOA/nph-KOAlogin}} (KOA).  We reduced the data from the KOA following the same steps as for our NIRSPEC data of HN~Peg~B (Section \ref{sec:NIRSPEC_HN_PegB}). For one of the comparison objects, SDSS 1021$-$03, the raw NIRSPEC N3-filter observations are not available in the KOA.  We downloaded the reduced spectrum for that object from the NIRSPEC Brown Dwarf Spectroscopic Survey (BDSS) online archive\footnote{\url{http://bdssarchive.org/}} \citep[][]{McLean_etal2003}. As the BDSS archive also contains a reduced spectrum of another one of our comparison objects, SDSS 1254$-$01, we compared the BDSS reduction of that object with ours. The two reductions of the NIRSPEC data for SDSS 1254$-$01 were nearly identical. Thus, we can say that the spectra of all five T2--T3 dwarfs were reduced following equivalent prescriptions.  The flux uncertainties of the spectra---plotted (except for SDSS 1021$-$03) along the bottom in the top left panel of Figure~\ref{fig:dwarfs_comparison}---are also comparable, so the observations attained a similar SNR of $\sim$50 at 1.25~$\micron$, except for HN~Peg~B which had SNR $\sim20$.

We show the $J$-band NIRSPEC spectrum of HN Peg~B and the four comparison T2--T3 dwarfs, along with their flux uncertainties (not available for SDSS 1021$-$03), in the top left panel of Figure~\ref{fig:dwarfs_comparison}. The top right panel of Figure~\ref{fig:dwarfs_comparison} zooms in on the longer of the two \ion{K}{1} doublets: at 1.2437~$\micron$ and 1.2529 $\mu$m. The shorter-wavelength \ion{K}{1} doublet at 1.175~$\micron$ has a lower SNR that hinders a comparison of the line strengths.

We measured the equivalent widths (EWs) of the \ion{K}{1} absorption lines at 1.2437~$\micron$ and 1.2529~$\mu$m in the NIRSPEC spectra of HN~Peg~B and the other four young and old T2--T3 dwarfs following the method of \citet{Allers-Liu2013}. The absorption line wavelength ranges were defined as shown by the shaded regions in the top right panel of Figure \ref{fig:dwarfs_comparison}.  In the bottom panel of Figure \ref{fig:dwarfs_comparison} we compare the EW measurements to the estimated surface gravities (Table \ref{tab:dwarfs_comparison}). The $J$-band K I doublet EWs of HN~Peg~B and the other comparison dwarfs cover a narrow range of $\approx$2~\AA\ (4.6--6.3~\AA\ and 6.8--9.1~\AA\ for the absorption lines at 1.2437~$\micron$ and 1.2529~$\micron$, respectively). There is no clear correlation between the \ion{K}{1} line strengths and surface gravity or age, although observations of additional targets could improve the robustness of this result. For comparison, the ranges of EWs of the 1.2529~$\micron$ \ion{K}{1} line in early-L ($\approx$4--10~\AA) or late-M ($\approx$2--6~\AA) dwarfs are 2--3 times wider over the same span of surface gravities, and the EWs show a clear positive correlation with gravity \citep{Allers-Liu2013,Martin_etal2017}.

Therefore, we conclude that the $J$-band \ion{K}{1} doublet absorption lines are inadequate as a surface gravity indicator for $4.3\lesssim\log g\lesssim 5.4$ early-T dwarfs, unlike for warmer M- and L-type dwarfs. It remains to be investigated whether this result extends to lower surface gravities and younger ages, as no such early-T dwarfs are presently known.

\defcitealias{Artigau_etal2006}{1}
\defcitealias{Pineda_etal2016}{2}
\defcitealias{Zuckerman_etal2006}{3}
\defcitealias{Gagne_etal2017}{4}
\defcitealias{Martin_etal2017}{5}
\defcitealias{Luhman_etal2007}{6}
\defcitealias{Zhou_etal2018}{8}
\defcitealias{Leggett_etal2000}{9}
\defcitealias{Burgasser_etal2003}{10}
\defcitealias{Burgasser_etal2006b}{11}
\defcitealias{Burgasser_etal2002}{12}
\defcitealias{Geballe_etal2002}{13}
\defcitealias{McLean_etal2003}{14}
\defcitealias{Stephens_etal2009}{15}
\defcitealias{Cushing_etal2008}{16}
\defcitealias{Filippazzo_etal2015}{17}
\defcitealias{Kirkpatrick_etal2008}{18}
\defcitealias{Vrba_etal2004}{19}
\defcitealias{Wielen1977}{20}
\defcitealias{Looper_etal2007}{21}
\defcitealias{Burgasser_etal2010b}{22}

\begin{table*}
\caption{Select properties of T2--T3 dwarfs observed with Keck/NIRSPEC in the N3 ($J$-band) filter.}
  \scriptsize
  \label{tab:dwarfs_comparison}
  \begin{threeparttable}
	\begin{tabularx}{\linewidth}{@{\extracolsep{-6pt}}lcccccccccccccccc}
    \toprule
	Name                             &  \multicolumn{2}{c}{SpT} & Age                    & $\log g$               & NIRSPEC-3     & \multicolumn{6}{c}{References}                                       \\
	\cline{2-3}
	\cline{7-12}
	                                 &  Optical     & Near-IR   &                        &                        & Obs. UT Date  & Discovery & SpT$_{\rm{Opt}}$ & SpT$_{\rm{NIR}}$ & Age & $\log g$ & Data \\
	                                 &              &           & (Gyr)                  & (dex)                  &               &           &                  &                  &     &          &      \\
    \midrule                                                                                            
	SIMP J013656.6+093347            & T2           & T2.5      & $\sim0.20\pm0.05$    & 4.31$\pm0.03$          & 2009 Nov 8    & \citetalias{Artigau_etal2006}  & \citetalias{Pineda_etal2016}                                      & \citetalias{Artigau_etal2006}                                                                                                & \citetalias{Zuckerman_etal2006},\citetalias{Gagne_etal2017}  & \citetalias{Gagne_etal2017}      & \citetalias{Martin_etal2017} \\
	HN~Peg~B (J21442847+1446077)$^a$ & \nodata      & T2.5      & $0.30_{-0.17}^{+0.28}$ & $4.66_{-0.25}^{+0.20}$ & 2008 Jul 8    & \citetalias{Luhman_etal2007}   & \nodata                                                           & \citetalias{Luhman_etal2007}                                                                                                 & 7                                                            & 7                                & \citetalias{Zhou_etal2018}   \\
	SDSS J125453.90$-$012247.4$^b$   & T2.0         & T2.0      & 0.1--2, $\sim$0.3      & 5.02$\pm$0.47          & 2001 Mar 7    & \citetalias{Leggett_etal2000}  & \citetalias{Burgasser_etal2003}                                   & \citetalias{Burgasser_etal2006b},\citetalias{Burgasser_etal2002},\citetalias{Geballe_etal2002},\citetalias{McLean_etal2003}  & \citetalias{Stephens_etal2009},\citetalias{Cushing_etal2008} & \citetalias{Filippazzo_etal2015} & \citetalias{McLean_etal2003} \\
	SDSS J102109.69$-$030420.1$^c$   & T2, T3.5$^d$ & T3.0$^d$  & $\gtrsim$1             & 5.1--5.5$^e$           & 2001 Jun 11   & \citetalias{Leggett_etal2000}  & \citetalias{Burgasser_etal2003},\citetalias{Kirkpatrick_etal2008} & \citetalias{Burgasser_etal2002},\citetalias{Burgasser_etal2006b},\citetalias{Geballe_etal2002},\citetalias{McLean_etal2003}  & \citetalias{Vrba_etal2004},\citetalias{Wielen1977}           & \nodata                          & \citetalias{McLean_etal2003} \\
	2MASS J11061197+2754225$^f$      & \nodata      & T2.5,T2   & $\gtrsim$2             & 5.3--5.5$^e$           & 2009 Apr 7    & \citetalias{Looper_etal2007}   & \nodata                                                           & \citetalias{Looper_etal2007},\citetalias{Burgasser_etal2010b}                                                                & \citetalias{Looper_etal2007},\citetalias{Wielen1977}         & \nodata                          & \citetalias{Martin_etal2017} \\
    \bottomrule
	\end{tabularx}
	\begin{tablenotes}[para,flushleft]
	$^a$Ruled out as a binary system with projected separation $\gtrsim$1.6 au for a distance of 18.13 pc (from \textit{HST}/NICMOS F090M observations in program GO~11304; PI: Metchev).\\
	$^b$Ruled out as a binary system with projected separation $\gtrsim$0.6 au for a distance of 13.5 pc \citep{Burgasser_etal2006}.\\
	$^c$Binary system composed of a T0--T1 and a T5--T5.5 dwarf \citep{Burgasser_etal2006,Dupuy-Liu2012,Burgasser_etal2010b}. \\
	$^d$Composite spectral type. \\
	$^e$Expected $\log g$ for the specified age and $T_{\rm eff}$=1150--1300 K, considering various sets of evolutionary models \citep[][]{Saumon-Marley2008,Marley_etal2018,Phillips_etal2020}.\\
	$^f$Ruled out as a binary system with projected separation $\gtrsim$1 au for a distance of 15.5 pc \citep{Looper_etal2008}.\\
	{\bf References:} (1) \citet{Artigau_etal2006}; (2) \citet{Pineda_etal2016}: (3) \citet{Zuckerman_etal2006}; (4) \citet{Gagne_etal2017}; (5) \citet{Martin_etal2017}; (6) \citet{Luhman_etal2007}; (7) This work; (8) \citet{Zhou_etal2018}; (9) \citet{Leggett_etal2000}; (10) \citet{Burgasser_etal2003}; (11) \citet{Burgasser_etal2006b}; (12) \citet{Burgasser_etal2002}; (13) \citet{Geballe_etal2002}; (14) \citet{McLean_etal2003}; (15) \citet{Stephens_etal2009}; (16) \citet{Cushing_etal2008}; (17) \citet{Filippazzo_etal2015}; (18) \citet{Kirkpatrick_etal2008}; (19) \citet{Vrba_etal2004}; (20) \citet{Wielen1977}; (21) \citet{Looper_etal2007}; (22) \citet{Burgasser_etal2010b}.
	\end{tablenotes}
 \end{threeparttable}
\end{table*}

\subsection{Comparisons to Atmospheric Models}
\label{sec:model_fits}
To determine the fundamental parameters of HN~Peg~B, which at a spectral type of T2.5 is expected to have some dust in its atmosphere, we compare the individual spectra and the assembled SED to photospheric models that include condensates: by \citet[][henceforth SM08]{Saumon-Marley2008} and \citet[][BT-Settl]{Allard_etal2012}. We also consider a comparison to the more recent (but cloudless) \citet[][Sonora 2018]{Marley_etal2018} and \citet[][ATMO 2020]{Phillips_etal2020} models.  However, we find that the SM08 and BT-Settl models perform better, producing fits with reduced chi-square (Equation \ref{eq:chi2_r}) values about twice smaller: 7--15 for models with condensates vs.\ 20--35 for models without. While both models with and without condensates struggle with reproducing the CH$_4$ and CO absorption strengths, condensate-free models exhibit greater challenges in the fundamental transitions between 3--5 $\mu$m, and may also significantly over-estimate the strength of H$_2$O absorptions in the near-IR. We analyze the results from the condensate models below, and relegate the analysis of the condensate-free models to  Appendix \ref{sec:cloudless_model_fits}. Separately, \citet{Tremblin_etal2019} offer an analysis of the IRTF/SpeX spectrum of HN~Peg~B with a condensate-free atmospheric model built on an enhanced treatment of (adiabatic and diabatic) convective instability. Because this model does reproduce the 0.9--2.4~$\mu$m $R\sim100$ spectrum comparably well, we include a comparison with it in Section~\ref{sec:SpeX_model_comparisons}.

In the following we describe in detail the fitting of the SM08 and BT-Settl models to each of the observed spectra (Sections~\ref{sec:NIRSPEC_model_comparisons}--\ref{sec:IRS_model_comparisons}) and to the full SED (Section~\ref{sec:all_model_comparisons}). In Table~\ref{tab:best-fitting_model_spectra} and Section~\ref{sec:comparison_best-fitting_models} we list and analyze the spectroscopic data sets that produce the most reliable and self-consistent sets of physical parameters when fit by models.  We summarize this in Table~\ref{tab:best_model_param}. The reader is welcome to skip directly to Section~\ref{sec:comparison_best-fitting_models} for the summary of findings.

We convolved the model photospheres to the same resolution as the observed spectra and fit them to the data by minimizing the reduced chi-square, $\chi^2_r$, defined as:

\begin{equation}
\chi^2_r = \frac{1}{N-n_p}\sum_{i=1}^{N} \left(\frac{O_i-\alpha M_i}{\sigma_i}\right)^2.
\label{eq:chi2_r}
\end{equation}
Here $N$ is the number of fitted points (data pixels), $n_p$ is the number of fitted model parameters (so $N-n_p$ are the degrees of freedom), $M_i$  and $O_i$ are the flux densities of the model and of the observed spectra, respectively, $\sigma_i$ are the uncertainties of the observed flux densities, and $\alpha$ is an unknown geometric dilution factor equal to $(R/d)^2$, where $R$ is the object radius and $d$ is the distance to the object. The value of $\alpha$ that minimizes $\chi^2_r$ is determined by  setting $\tfrac{\partial \chi^2_r}{\partial \alpha}=0$, and is obtained as:

\begin{equation}
\alpha = \dfrac{\sum\limits_{i=1}^{N}\dfrac{O_iM_i}{\sigma_i^2}} {\sum\limits_{i=1}^{N}\dfrac{M_i^2}{\sigma_i^2}}.
\label{eq:alpha_chi2_r}
\end{equation}

When fitting SM08 models we used synthetic spectra covering the following parameter ranges: $800\le T_{\rm eff}/{\rm{K}} \le2000$, $4.0\le \log g \le 5.5$ (down to $\log g=3.0$ for $T_{\rm eff}=1000-1200$ K and $f_{\rm sed}=2$) and $1 \le f_{\rm sed}\le 4$ in steps of 100~K, 0.5 dex, and 1.0, respectively (same as in the provided model grid). The parameter $f_{\rm sed}$ is the sedimentation efficiency that indicates the size of atmospheric condensate particles. Larger $f_{\rm sed}$ values imply larger mean particle sizes for the cloud condensates that then fall out of the visible atmosphere faster. Atmospheres with small $f_{\rm sed}$ values are dusty and optically thicker. For the BT-Settl model fits we considered synthetic spectra with the following parameters: $500\le T_{\rm eff}/\rm{K}\le 2000$ and $2.0 \le \log g \le 5.5$ (3.0--5.0 for $T_{\rm eff}<1000$ K) in steps of 100~K or 50~K for some cases (mainly for cool atmospheres) and 0.5 dex, respectively (same as in the provided model grid). We considered atmospheres with solar metallicity from both sets of models, in agreement with the metallicity of HN~Peg~A \citep[$-0.01\pm0.03$;][]{Valenti-Fischer2005}. 

In the following model comparisons (Sections \ref{sec:NIRSPEC_model_comparisons}--\ref{sec:comparison_best-fitting_models}) we consider the five best-fit models to each of our different data sets.  In doing so we assess the approximate systematic uncertainties in the fundamental parameters arising from the models, and compare the performance of the two families of models with condensates.

\subsubsection{Model Comparisons to the Keck NIRSPEC Spectrum}
\label{sec:NIRSPEC_model_comparisons}
\begin{figure*}
	\centering
	\begin{tabular}{cc}
		\subfloat{\includegraphics[width=.50\linewidth]{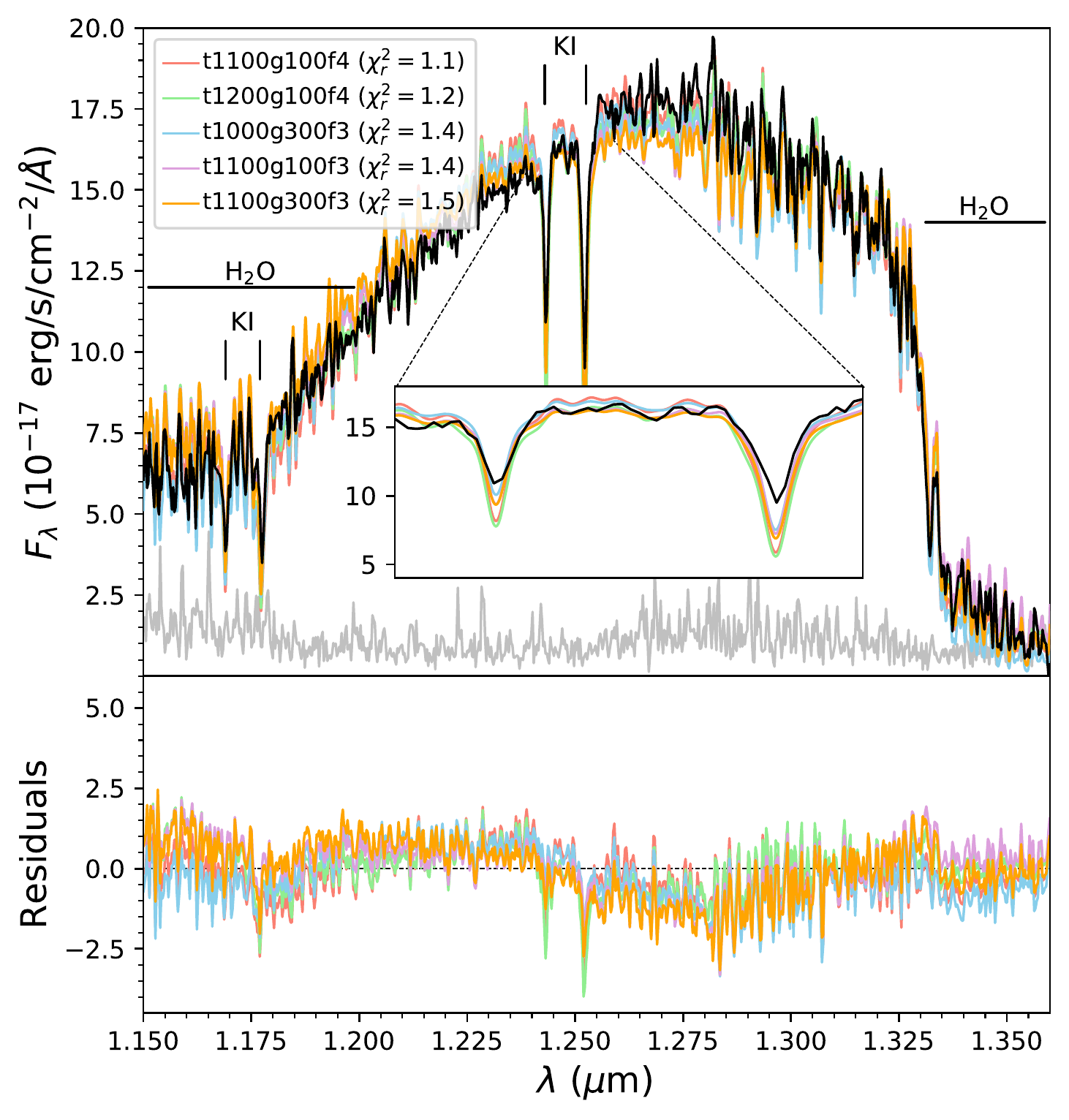}} & 
		\subfloat{\includegraphics[width=.50\linewidth]{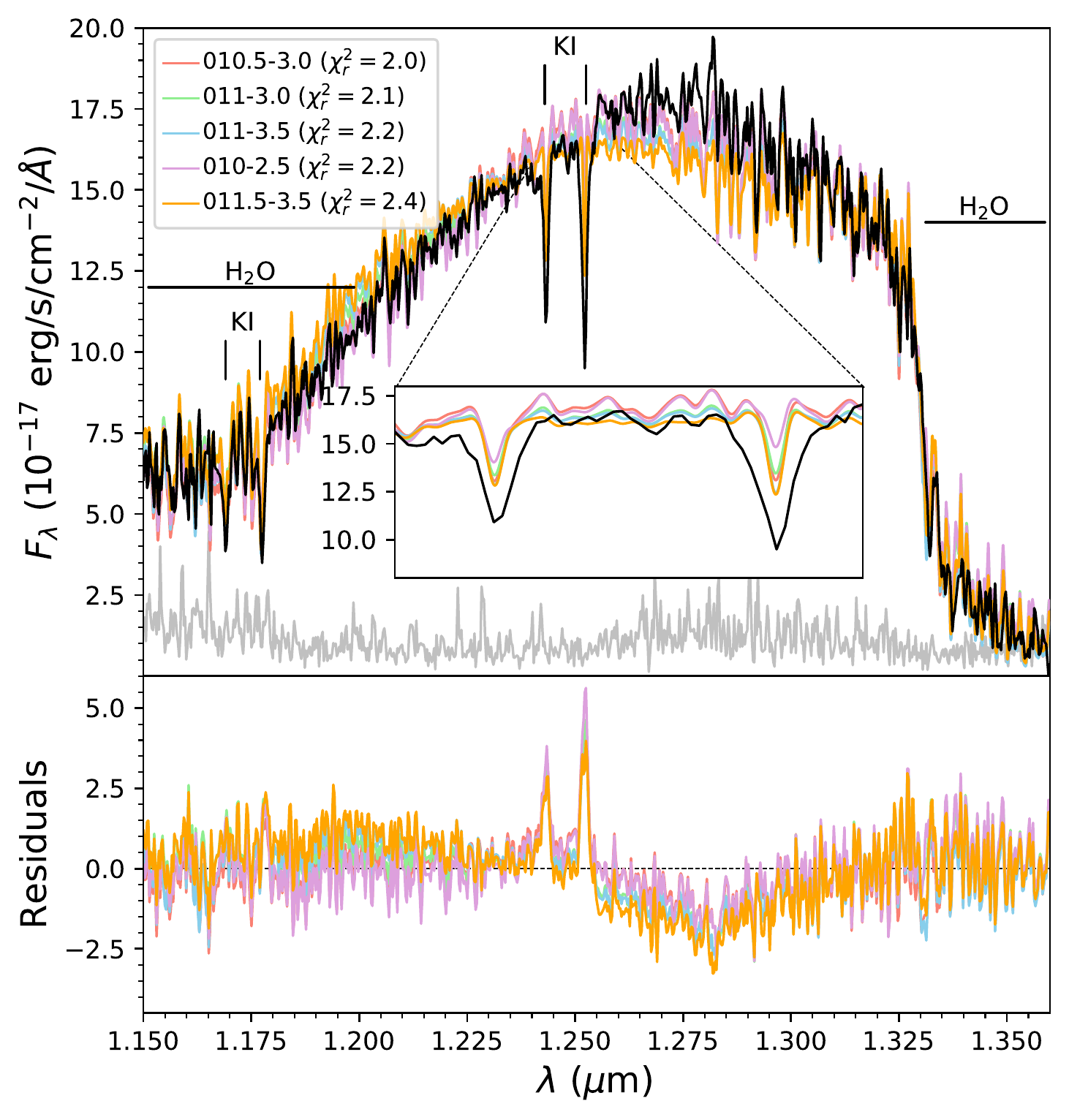}} 
	\end{tabular}
	\caption{{\bf Upper panels:} Photospheric fits (colored curves) to the overall 1.15--1.35 $\mu$m $R\approx2300$ NIRSPEC spectrum of HN~Peg~B (black) with SM08 (left panel) and BT-Settl (right panel) models.  The legends show the parameters of the synthetic spectra in the native model nomenclature ($T_{\rm eff}$ in K, $g$ in m~s$^{-2}$ and $f_{\rm sed}$ for SM08; $T_{\rm eff}$ in K/100 and $\log g$ (with $g$ in cm~s$^{-2}$) for BT-Settl) together with the $\chi^2_r$ values of the fits.  The uncertainties of the NIRSPEC spectrum are shown by the gray curve along the bottom of the plots.  The main spectral features are indicated. {\bf Bottom panels:} Residuals between the best model fits and the NIRSPEC spectrum (same units and scaling as in upper panels).}
	\label{fig:model_comparisons_NIRSPEC_all_best}
\end{figure*}

In Figure~\ref{fig:model_comparisons_NIRSPEC_all_best} we show the five best-fitting SM08 (left panel) and BT-Settl (right panel) model spectra to the HN~Peg~B NIRSPEC spectrum. The fits were done by minimizing $\chi^2_r$ over the whole 1.15--1.36 $\mu$m wavelength range.

\begin{figure*}
	\centering
	\begin{tabular}{cc}
		\subfloat{\includegraphics[width=.50\linewidth]{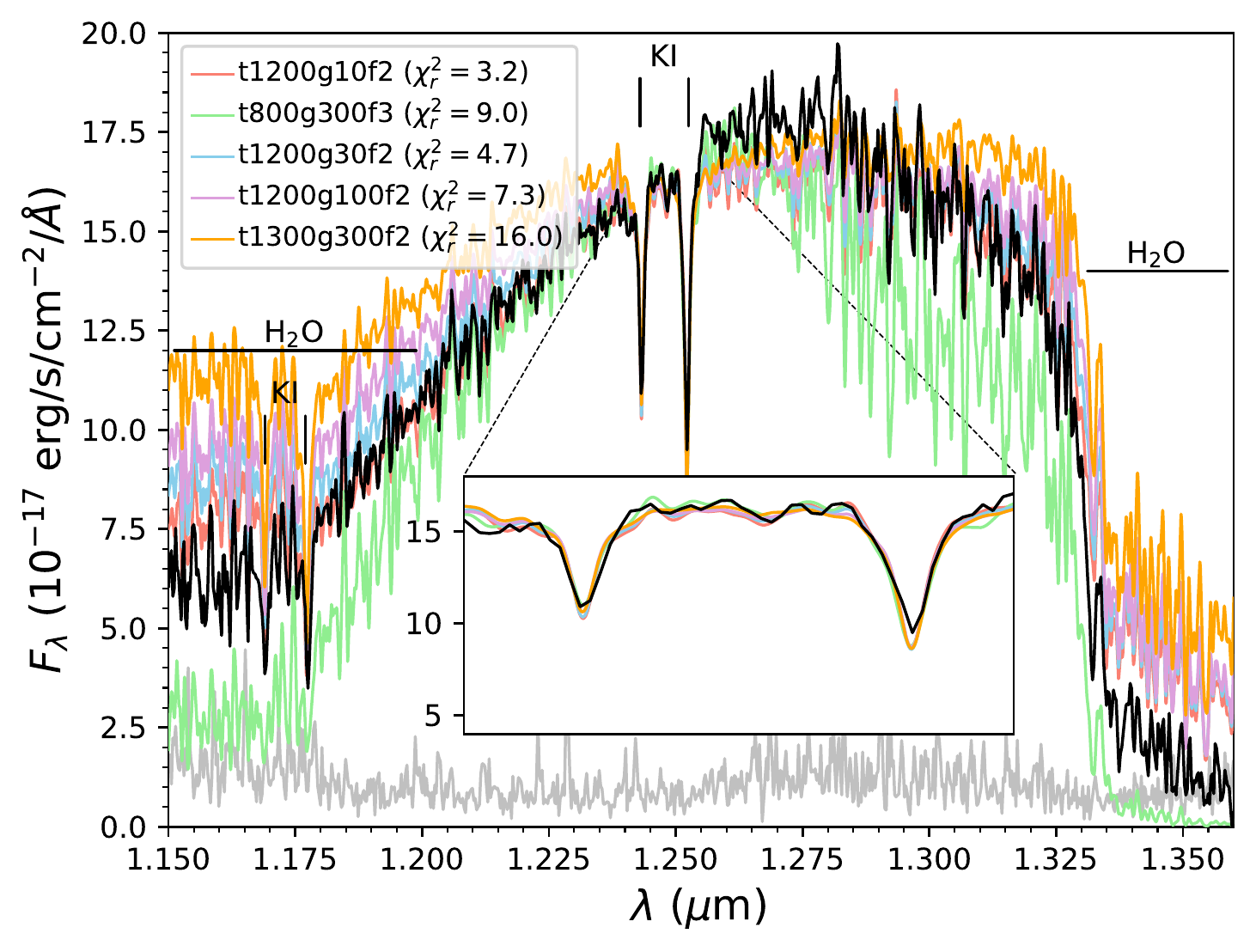}} & 
		\subfloat{\includegraphics[width=.50\linewidth]{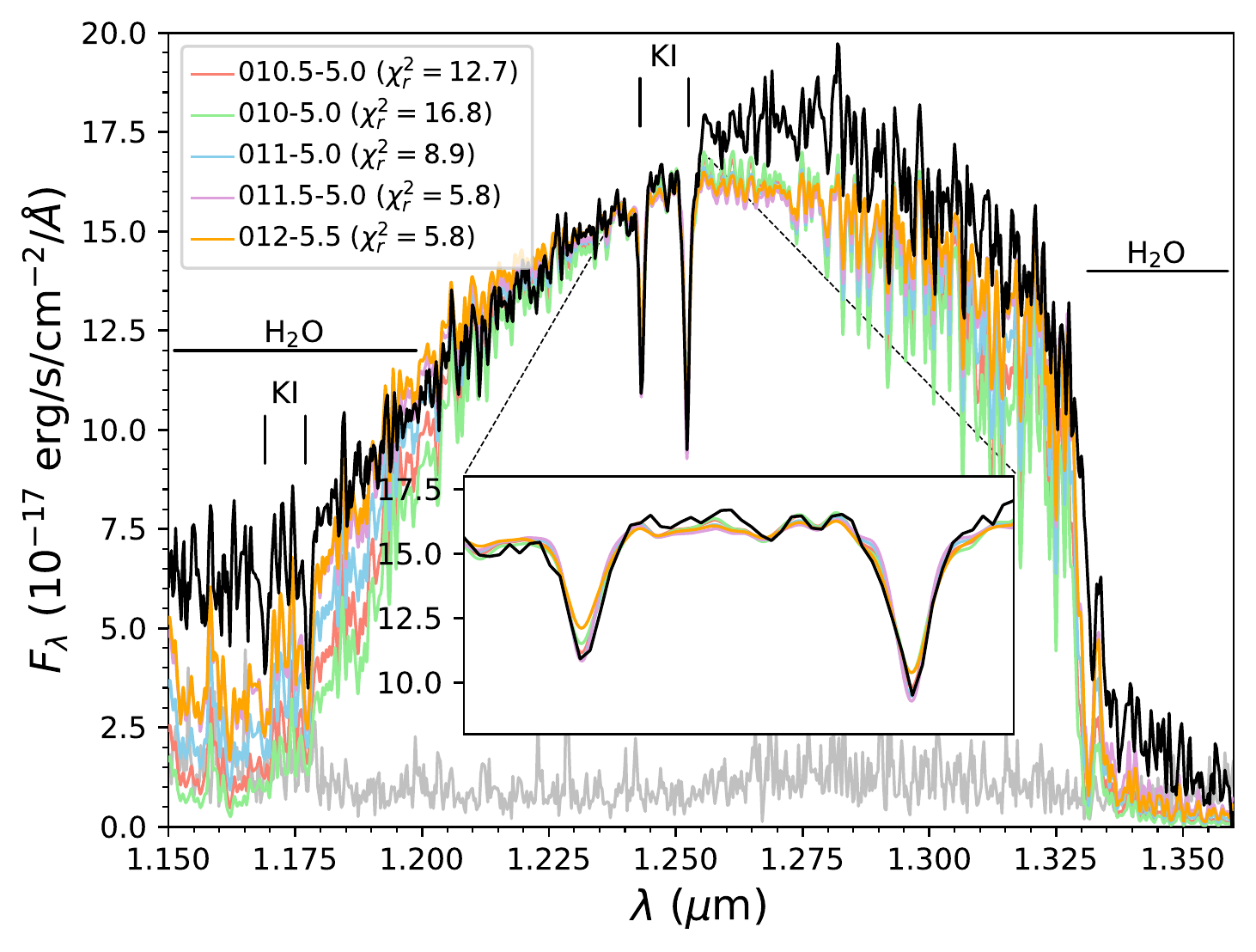}} 
	\end{tabular}
	\caption{Fits to the \ion{K}{1} doublet at 1.25 $\mu$m in the NIRSPEC spectrum of HN~Peg~B with models from SM08 (left panel) and BT-Settl (right panel). The legends have similar nomenclature as in Figure~\ref{fig:model_comparisons_NIRSPEC_all_best}. The $\chi^2_r$ values are for the fits of the synthetic spectra to the observed spectra over the entire NIRSPEC wavelength range.}
	\label{fig:KI_test}
\end{figure*}

The best-fitting SM08 model spectra represent the overall NIRSPEC spectrum well, with $\chi^2_r$ between 1.1 and 1.5.  Based on the ensemble of these fits, HN~Peg~B has 1000~K $\le T_{\rm eff} \le 1200$~K, $4.0\le \log g\le 4.5$ and $3 \le f_{\rm sed} \le 4$ (Table \ref{tab:best_model_param}). The best-fitting BT-Settl photospheric models do not match the NIRSPEC spectrum nearly as well, with $\chi^2_r$ values between 2.0 and 2.4.  The inferred $T_{\rm eff}$ from BT-Settl is similar (1000~K $\le T_{\rm eff} \le 1150$~K), but the surface gravity is an order of magnitude lower: $2.5 \le \log g \le 3.5$ (Table \ref{tab:best_model_param}). 

Both sets of models (SM08 and BT-Settl) fail to accurately reproduce the depth of the \ion{K}{1} doublet at 1.25 $\mu$m (Figure~\ref{fig:model_comparisons_NIRSPEC_all_best}).  The SM08 models predict significantly stronger \ion{K}{1} absorption, while the BT-Settl models predict a significantly weaker one.  The discrepancies are similar in the shorter-wavelength \ion{K}{1} doublet at 1.175 $\mu$m, although less pronounced because of the lower signal-to-noise ratio of the data.  Among the five best-fit photospheres from each of the SM08 and BT-Settl models, the ones with higher gravity match the depths of the \ion{K}{1} lines better: $g=300$~m~s$^{-2}$ (i.e., $\log g=4.5$) for SM08 and $\log g=3.5$ for BT-Settl.  In the SM08 case that corresponds to the weakest of the modelled \ion{K}{1} absorption strengths, whereas in the BT-Settl case that corresponds to the strongest modelled absorption strengths.

To better match the strength of the \ion{K}{1} absorption the SM08 models require either a lower sedimentation efficiency ($f_{\rm sed}=2$) with a potentially lower surface gravity ($2\le\log g\le4.5$) or a significantly cooler ($T_{\rm eff}=800$~K) atmosphere (Figure~\ref{fig:KI_test}, left panel).  The BT-Settl models require very high surface gravity ($5\le\log g\le5.5$; Figure~\ref{fig:KI_test}, right panel), as for $>$1~Gyr-old field T dwarfs \citep[e.g.,][]{Filippazzo_etal2015}. In both cases, when trying to match the \ion{K}{1} strengths, the overall fits to the $J$-band NIRSPEC spectrum become significantly poorer ($\chi^2_r>3$).  As we already noted in Section~\ref{sec:results:ki}, the \ion{K}{1} absorption strength is not sensitive to gravity in T2--T3 dwarfs.  Hence, this is an area in the atmospheric modeling that could benefit from further study.

\subsubsection{Model Comparisons to the IRTF SpeX Spectrum}
\label{sec:SpeX_model_comparisons}
\begin{figure*}
	\centering
	\begin{tabular}{cc}
		\subfloat{\includegraphics[width=.50\linewidth]{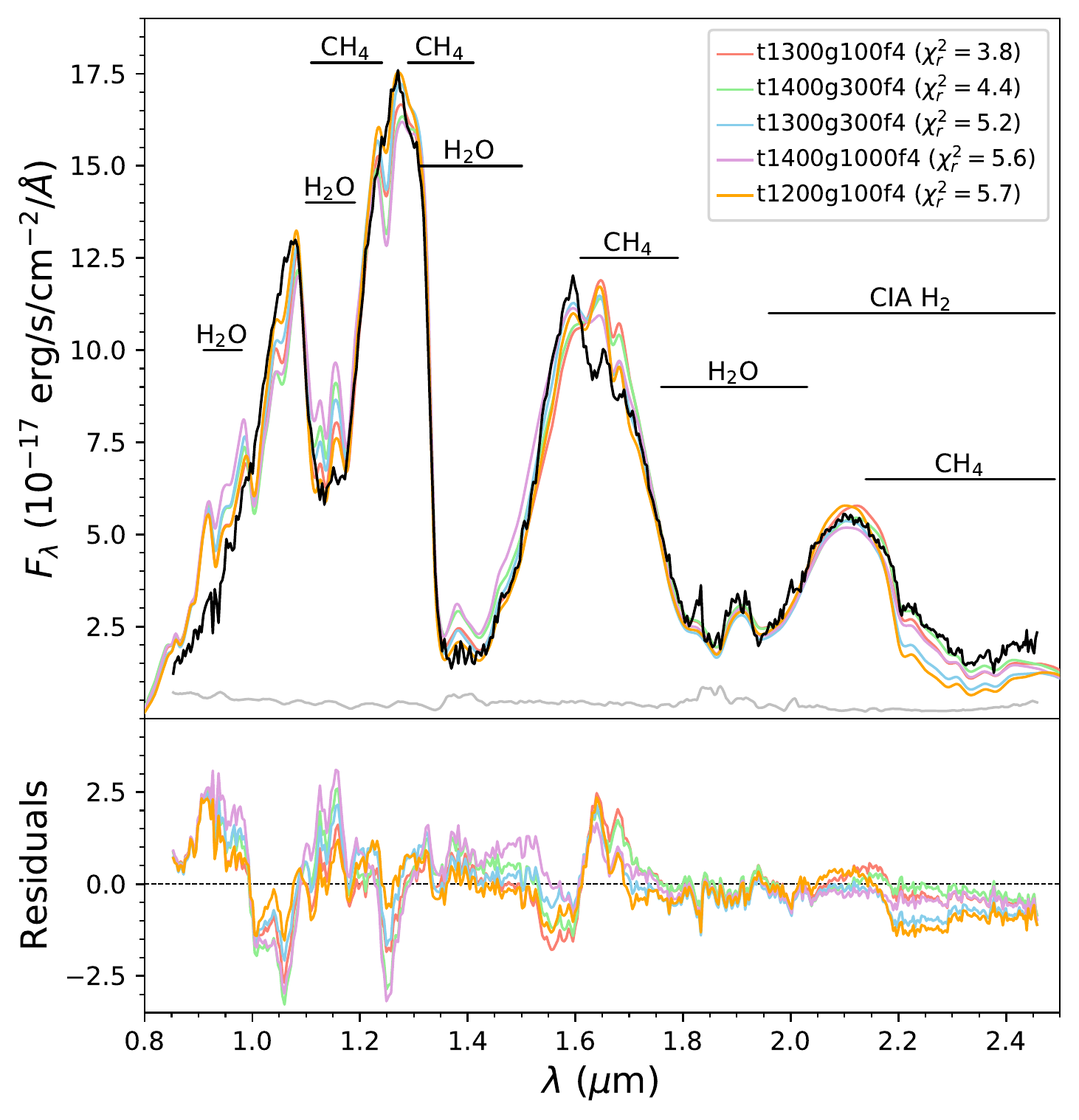}} & 
		\subfloat{\includegraphics[width=.50\linewidth]{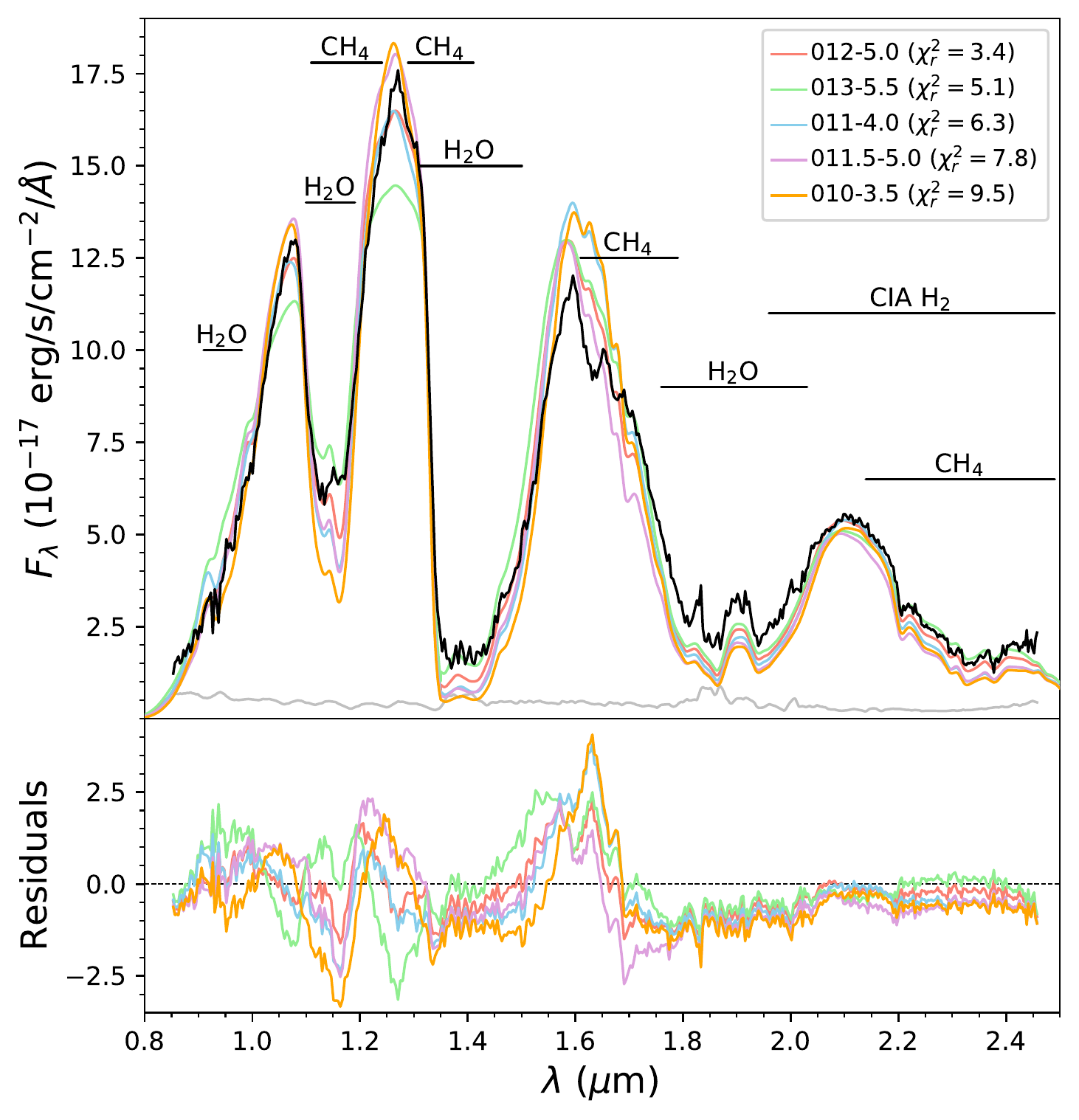}} 
	\end{tabular}
	\caption{{\bf Upper panels:} Best SM08 (left panel) and BT-Settl (right panel) model spectra fits to the $R\sim100$ SpeX spectrum of HN~Peg~B. The uncertainties of the SpeX spectrum are shown by the gray curve along the bottom of the plots. Legends have similar nomenclature as in Figure~\ref{fig:model_comparisons_NIRSPEC_all_best}. Some key molecular features are indicated. {\bf Bottom panels:} Residuals between the best model fits and the SpeX spectrum (same units and scaling as in the upper panels).}
	\label{fig:model_comparisons_SpeX_all_best}
\end{figure*}

In Figure \ref{fig:model_comparisons_SpeX_all_best} we show the best-fitting SM08 (left panel) and BT-Settl (right panel) model spectra to the SpeX spectrum of HN~Peg~B over 0.8--2.5 $\mu$m. Overall, both sets of models adequately reproduce the data, with $\chi^2_r$ values ranging between 3.4 and 9.5 for the five best-fit photospheres from each set of models.  The greatest challenges lie in reproducing the strength of the methane absorption in the $H$ and $K$ bands (both sets of models), the red optical slope (SM08 models), and the strength of the water bands (BT-Settl models).

As summarized in Table \ref{tab:best_model_param}, the parameters of the best-fitting synthetic spectra are:  $1200\le T_{\rm eff}/{\rm K}\le1400$, $4.0\le \log g\le5.0$ and $f_{\rm sed}=4$ for the SM08 models and $1000 \le T_{\rm eff}/{\rm K}\le 1300$, $3.5\le \log g\le 5.5$ for the BT-Settl models. 

We indicated at the beginning of Section~\ref{sec:model_fits} that atmospheric models with condensates reproduce the spectrum of HN~Peg~B better than models that are condensate-free.  A notable exception is the analysis of the same HN~Peg~B IRTF/SpeX spectrum by \citet{Tremblin_etal2019} with a custom condensate-free model based on a general diabatic convective instability treatment.  \citet[][and references therein]{Tremblin_etal2019} argue that disequilibrium chemistry of carbon-bearing molecules can itself trigger convective instability, similarly to thermohaline convection in Earth's oceans or to fingering convection in stellar atmospheres. These instabilities can significantly reduce the temperature gradient in substellar atmospheres, which can then explain several phenomena observed in brown dwarfs commonly attributed to condensate clouds, such as the rapid spectroscopic changes at the L/T transition \citep{Tremblin_etal2016} or the extremely red colors of low-gravity young dwarfs \citep{Tremblin_etal2017}. This treatment offers an alternate account of ultra-cool atmospheres that does not invoke condensates.

We compare the fit to the HN Peg B SpeX spectrum in Figure 5 of \citet{Tremblin_etal2019} to our own fits with the cloudy SM08 and BT-Settl models (Figure \ref{fig:model_comparisons_SpeX_all_best}). The three sets of models produce similarly adequate fits. A goodness of fit metric is not calculated in the \citet{Tremblin_etal2019} analysis for an exact comparison. Overall, there is a good match to the entire SpeX spectrum in all cases, with the main challenge being the methane absorption strength in the $H$ band: the best SM08 or BT-Settl model photospheres underestimate the methane absorption at $H$ band while the \citet{Tremblin_etal2019} model overestimates it. For the best-fit SM08 models, this could be a function of their higher effective temperatures (1200 K $\leq T_{\rm eff} \leq$ 1400~K) compared to the $T_{\rm eff}=1150$ K best-fit \citet{Tremblin_etal2019} model.  For BT-Settl (1000 K $\leq T_{\rm eff} \leq$ 1300~K) the reason is unclear.

The \citet{Tremblin_etal2019} general diabatic convection model is promising for analyzing the atmospheres of early-T dwarfs, such as HN~Peg~B.  However, a publicly available version of the model is yet to be released.  We note that although the \citet{Tremblin_etal2019} models and the public \citet{Phillips_etal2020} ATMO~2020 models were generated with the same code (ATMO), our ATMO~2020 model comparisons to the HN~Peg~B spectra (see Appendix; Figure \ref{fig:SED_test}) do not produce results that are as good as in \citet{Tremblin_etal2019}.  The likely reason is the improved treatment of convective instability in \citet{Tremblin_etal2019} that incorporates both adiabatic and diabatic terms.

\subsubsection{Model Comparisons to the Gemini NIRI Spectrum}
\label{sec:NIRI_model_comparisons}
\begin{figure*}
	\centering
	\begin{tabular}{cc}
		\subfloat{\includegraphics[width=.50\linewidth]{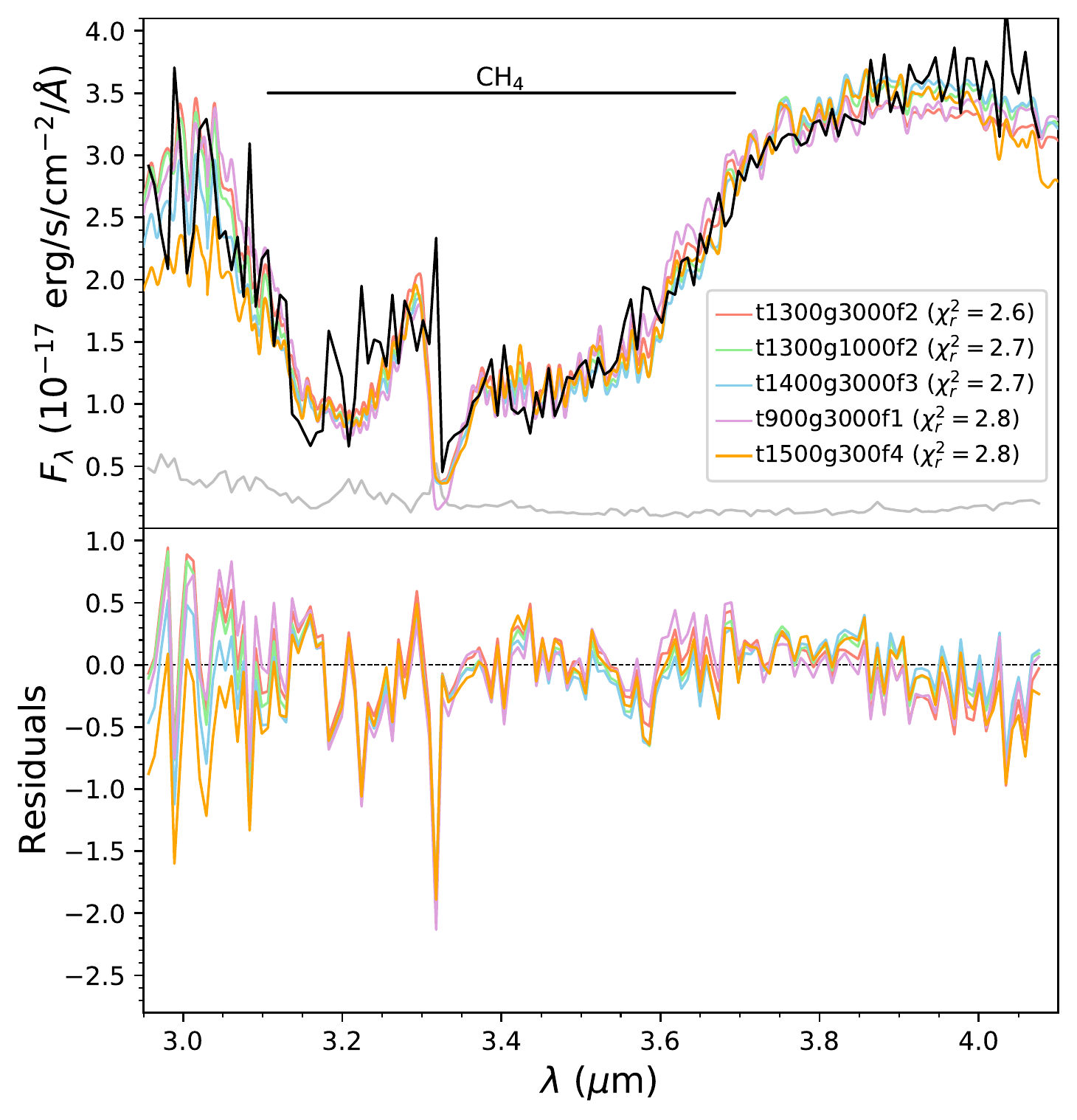}} & 
		\subfloat{\includegraphics[width=.50\linewidth]{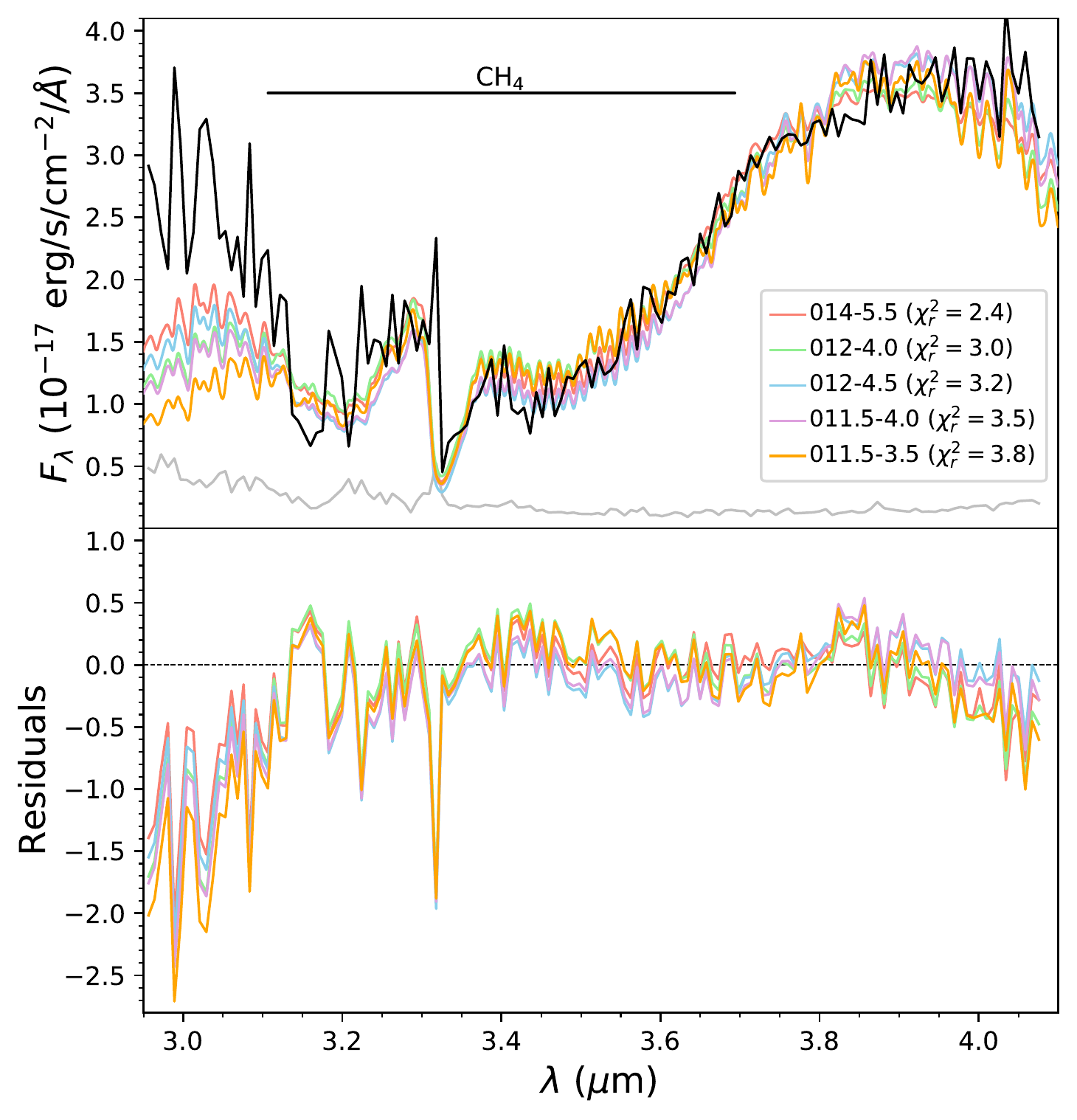}} 
	\end{tabular}
	\caption{{\bf Upper panels:} Best SM08 (left panel) and BT-Settl (right panel) model spectra fits to the $R\approx400$ NIRI spectrum of HN~Peg~B. The uncertainties of the NIRI spectrum are shown by the gray curve along the bottom of the plots. Legends have similar nomenclature as in Figure~\ref{fig:model_comparisons_NIRSPEC_all_best}. The main CH$_4$ molecular band is indicated. {\bf Bottom panels:} Residuals between the best model fits and the NIRI spectrum (same units and scaling as in the upper panels).}
	\label{fig:model_comparisons_NIRI_all_best}
\end{figure*}

The best-fitting SM08 and BT-Settl model spectra to the NIRI spectrum of HN~Peg~B are shown in the left and right panels of Figure \ref{fig:model_comparisons_NIRI_all_best}, respectively.  The NIRI spectrum is dominated by the CH$_4$ absorption at 3.3 $\mu$m, although also includes coverage of the bright flux peak at $\sim4\ \mu$m. The spectral shape is well reproduced by the two sets of synthetic spectra, with $\chi^2_r$ in the range between 2.4 and 3.8 for the five best-fit models in each case. The best-fitting BT-Settl synthetic spectra under-predict the flux at the blue end of the CH$_4$ absorption.  However, that is also the wavelength region with the lowest SNR in the NIRI spectrum, so it does not drive the quality of the overall fit.

Overall, the narrow wavelength range and the relatively low spectral resolution of the NIRI spectrum are not strongly diagnostic of the physical parameters of HN~Peg~B. As summarized in Table \ref{tab:best_model_param}, the parameters of the best-fitting synthetic spectra are: $900\le T_{\rm eff}/{\rm K}\le1500$, $4.5\le\log g\le5.5$, and $1\le f_{\rm sed}\le 4$ for the SM08 models, and  $1150\le T_{\rm eff}/{\rm K}\le1400$, $3.5\le\log g\le5.5$ for the BT-Settl models.

\subsubsection{Model Comparisons to the {\it Spitzer} IRS Spectrum}
\label{sec:IRS_model_comparisons}
\begin{figure*}
	\centering
	\begin{tabular}{cc}
		\subfloat{\includegraphics[width=.50\linewidth]{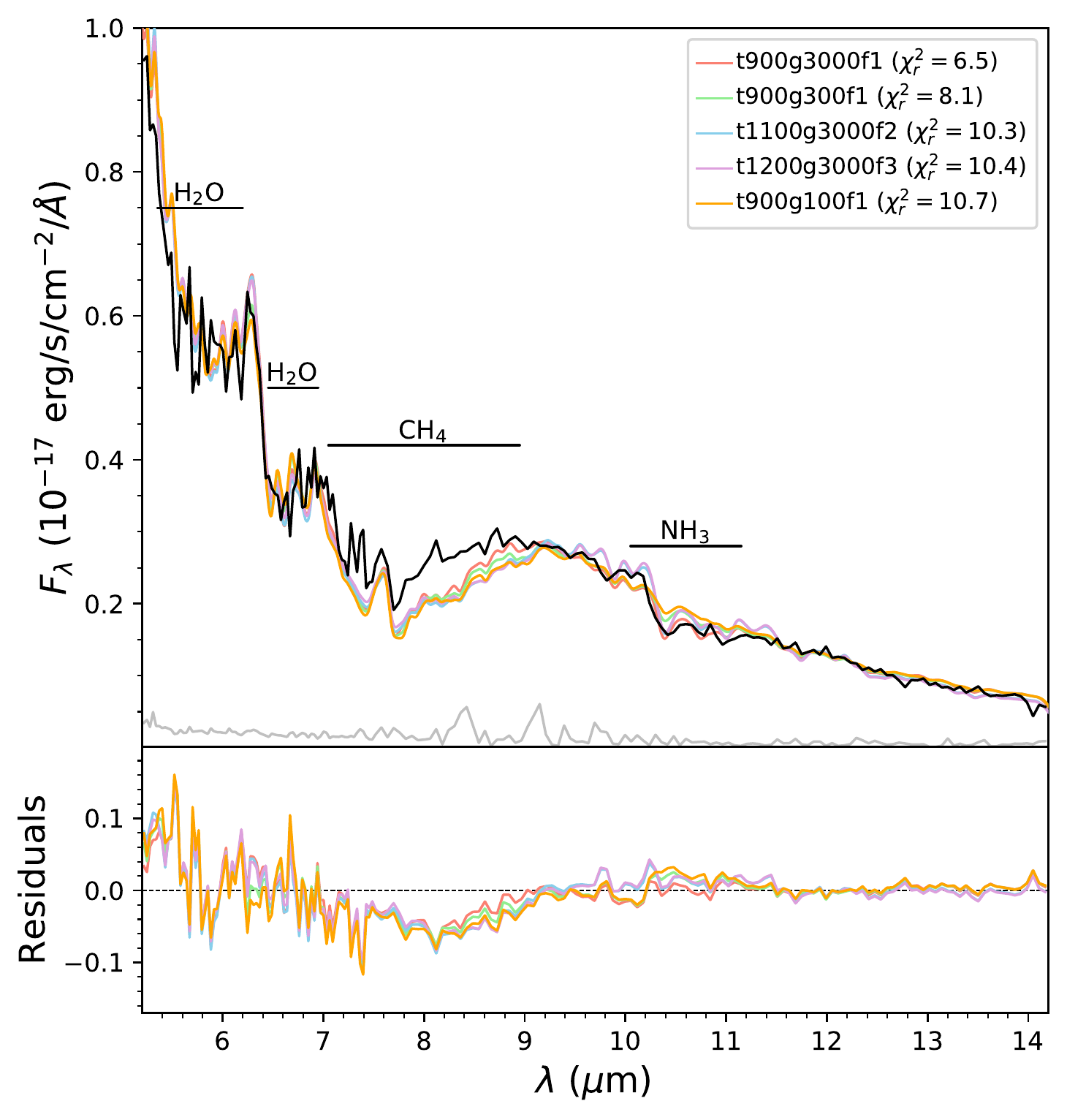}} & 
		\subfloat{\includegraphics[width=.50\linewidth]{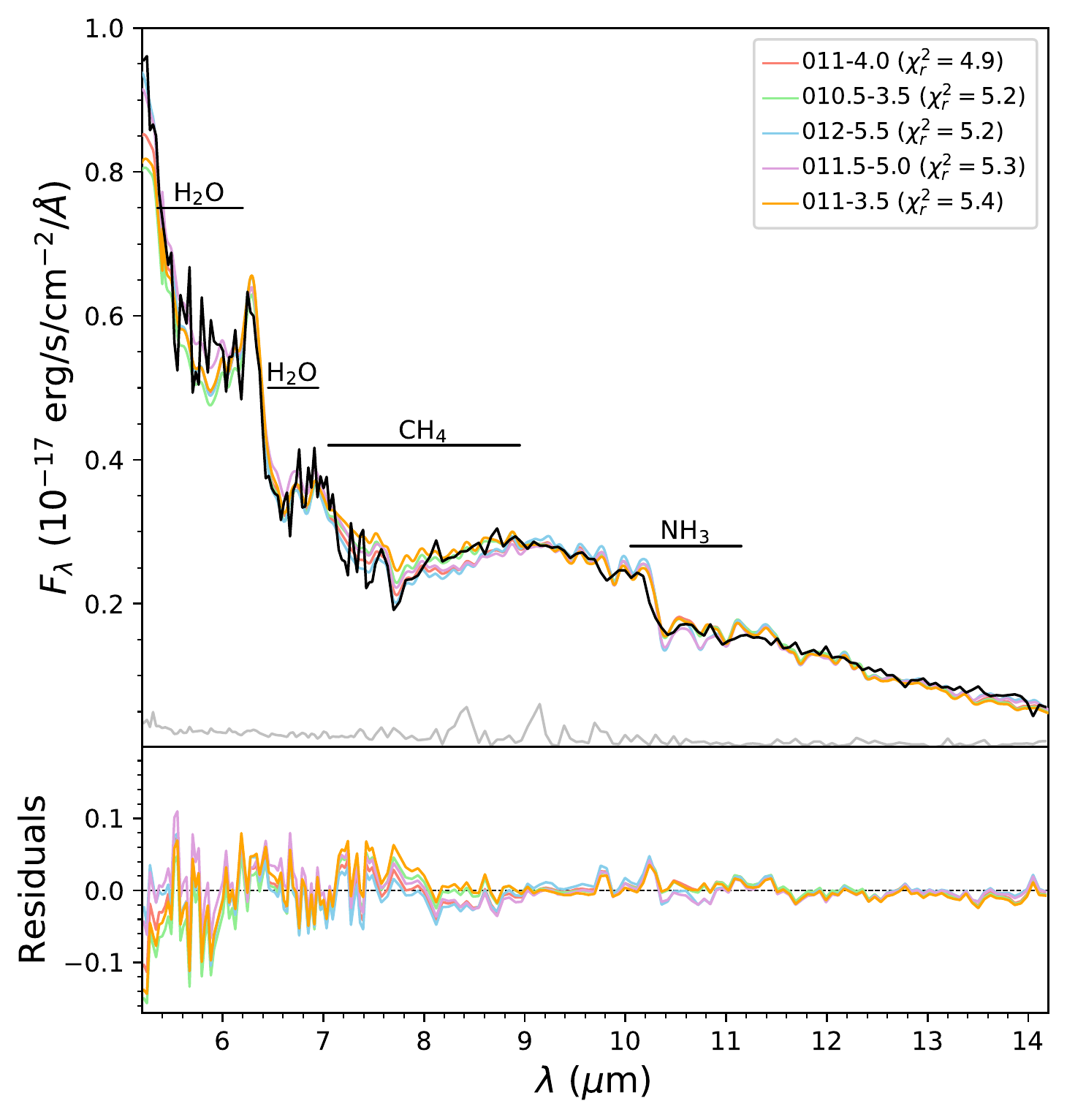}} 
	\end{tabular}
	\caption{{\bf Upper panels:} Best SM08 (left panel) and BT-Settl (right panel) model spectra fits to the $R\approx 60-130$ IRS spectrum of HN~Peg~B. The uncertainties of the IRS spectrum are shown by the gray curve along the bottom of the plots. Legends have similar nomenclature as in Figure~\ref{fig:model_comparisons_NIRSPEC_all_best}. Some key molecular features are indicated. {\bf Bottom panels:} Residuals between the best model fits and the IRS spectrum (same units and scaling as in the upper panels).}
	\label{fig:model_comparisons_IRS_all_best}
\end{figure*}

In Figure \ref{fig:model_comparisons_IRS_all_best} we show the best SM08 (left panel) and BT-Settl (right panel) model spectra fits to the HN~Peg~B IRS spectrum.  The five best-fitting BT-Settl model  photospheres reproduce the overall shape of the IRS spectrum better than the corresponding five best-fitting SM08 model photospheres, with $\chi^2_r$ values that are 25\%--50\% smaller.  The main challenge for the SM08 models is the strength of the CH$_4$ absorption, as in the case of the low resolution 0.8--2.5$\mu$m SpeX spectrum (Section~\ref{sec:SpeX_model_comparisons}).

The best-fitting SM08 spectra indicate similarly broad parameter ranges as for the NIRI spectrum: $900\le T_{\rm eff}/{\rm K}\le 1200$, $4.0\le \log g\le 5.5$ and $1\le f_{\rm sed}\le 3$ (Section~\ref{sec:NIRI_model_comparisons}; Table~\ref{tab:best_model_param}). The BT-Settl models point to a narrower temperature range ($1050\le T_{\rm eff}/{\rm K}\le 1200$) and a similarly broad surface gravity range ($3.5\le \log g\le 5.5$).

\subsubsection{Model Comparisons to All Spectra and Photometry: Challenges in the Red Optical and in CH$_4$/CO Chemical Disequilibrium}
\label{sec:all_model_comparisons}
\begin{figure*}
	\centering
	\begin{tabular}{cc}
		\subfloat{\includegraphics[width=.50\linewidth]{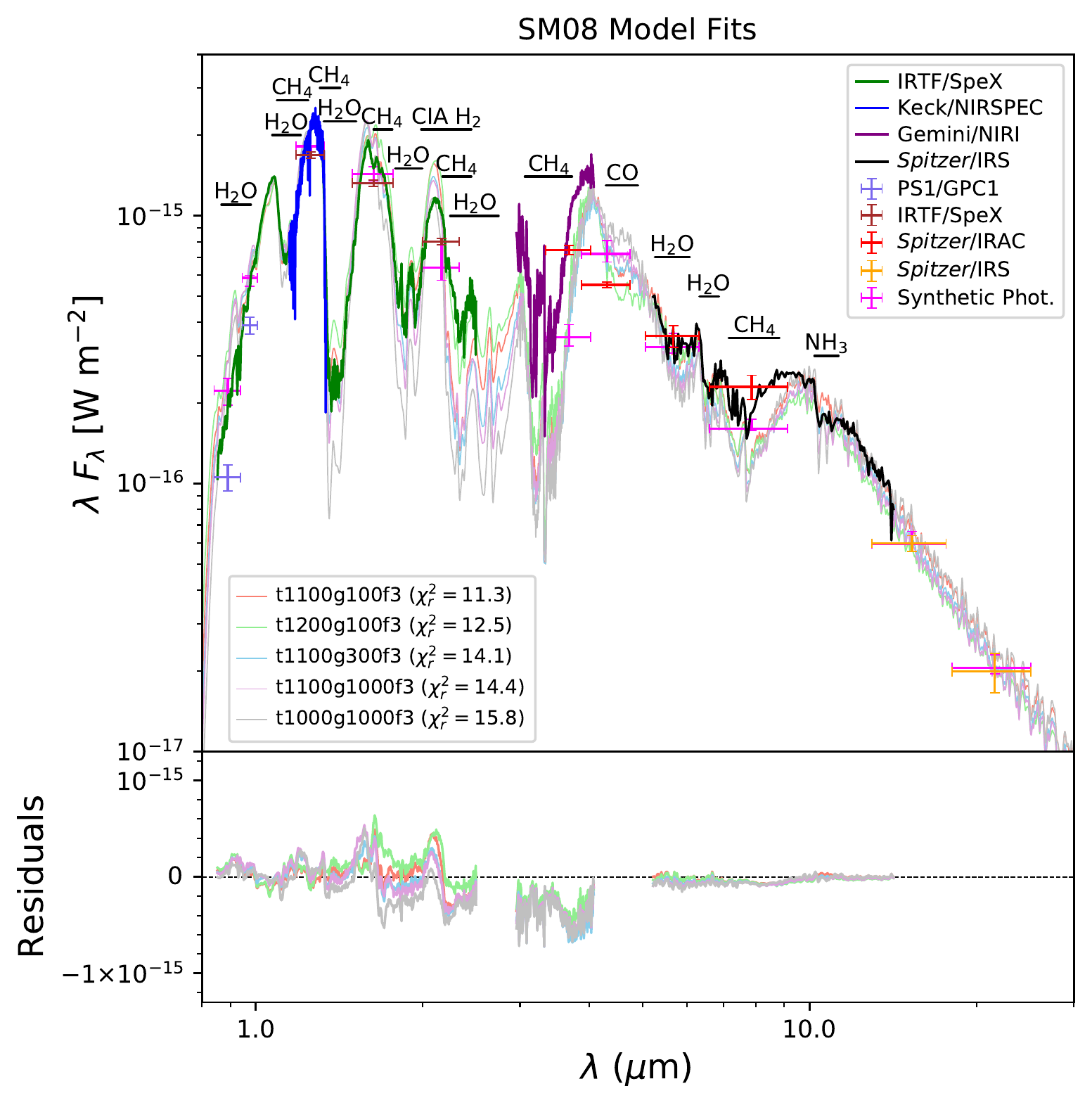}} &
		\subfloat{\includegraphics[width=.50\linewidth]{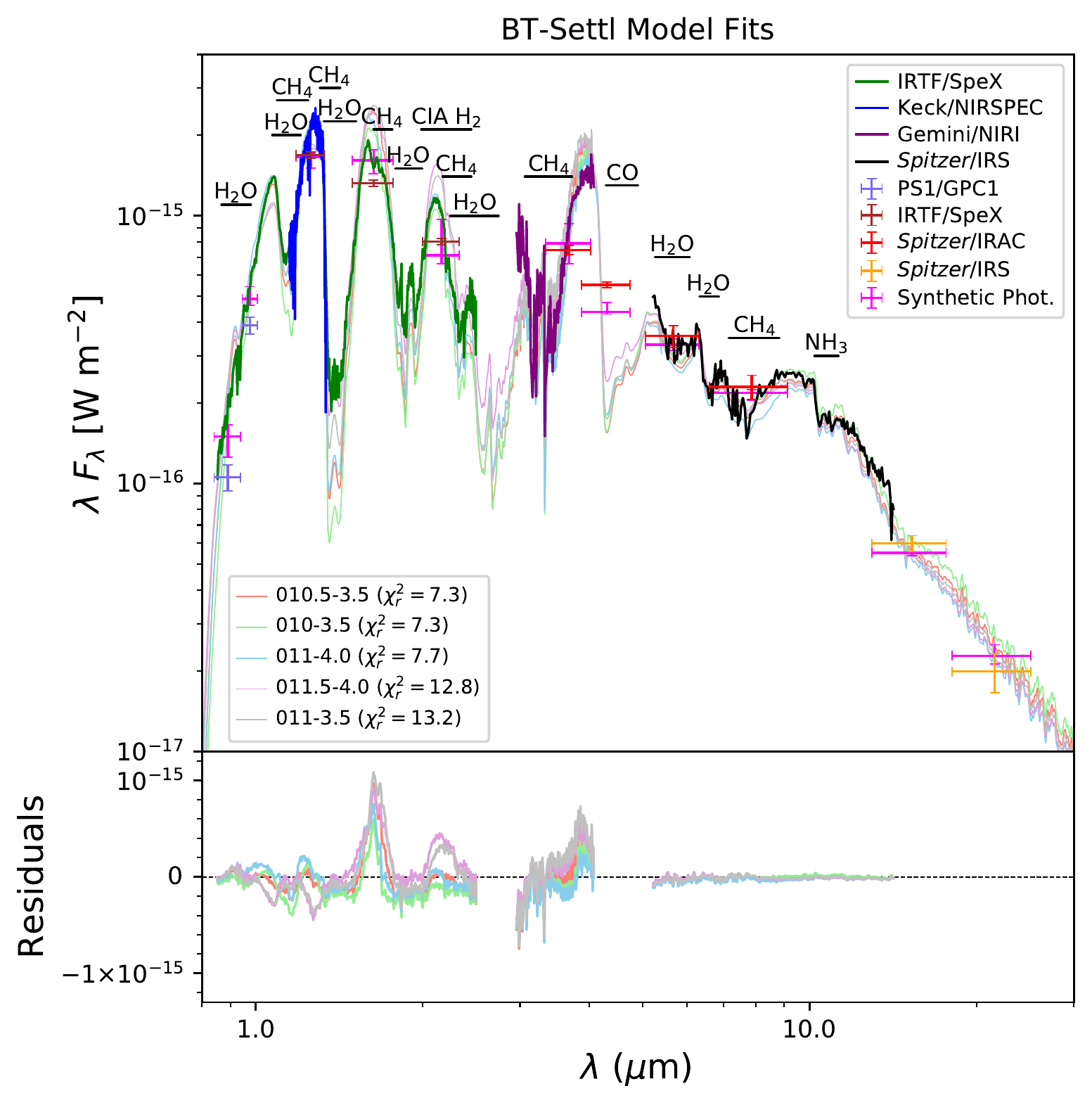}} 
	\end{tabular}
	\caption{{\bf Upper panels:} SED of HN~Peg~B using the assembled flux-calibrated near- and mid-IR spectrophotometry, with measurement instruments indicated in the top-right legends. The best-fitting SM08 (left panel) and BT-Settl (right panel) model spectra to the spectrophotometry are shown by the curves indicated in the bottom legends, which have similar nomenclature as in Figure~\ref{fig:model_comparisons_NIRSPEC_all_best}. The horizontal lines correspond to the passband width of each filter centered at its effective wavelength (Table \ref{tab:HN_PegB_phot}). The main spectral features are indicated. {\bf Bottom panels:} Residuals between the best model fits and the data (same units as in the upper panels). \\ (All spectra are available in the online journal.)}
	\label{fig:SED}
\end{figure*}

In addition to the model comparisons to the individual spectra presented in Sections~\ref{sec:NIRSPEC_model_comparisons}--\ref{sec:IRS_model_comparisons}, we also fit SM08 and BT-Settl synthetic photospheres to all spectra and photometry (hereafter referred as full SED) of HN~Peg~B simultaneously. Figure \ref{fig:SED} shows one of the most spectrophotometrically complete SED of an L/T transition dwarf, ranging from 0.8 $\mu$m to 26.3 $\mu$m.  The presented spectrophotometry encompasses 98\% of the total bolometric luminosity of HN~Peg~B.

The model fits reproduce the spectra and the SED fairly well, even if the $\chi^2_r$ values are relatively high.  The three best-fitting BT-Settl models have $7<\chi^2_r<8$, while all other models have $\chi^2_r>10$.  The greatest shortcoming of the SM08 models seems to be over-predicting the strength of the CH$_4$ absorption at 3--4~$\mu$m and 7--9~$\mu$m.  Notably, the entire 3.0--4.1 $\mu$m NIRI spectrum, which is dominated by the methane fundamental band at 3.3~$\mu$m, lies systematically above all five of the best-fitting SM08 models.  This was not evident in our fitting of the NIRI spectrum alone (Section~\ref{sec:NIRI_model_comparisons}) because we allow scaling as a free parameter for the models to match the individual spectra (Equation~\ref{eq:chi2_r}).

As summarized in Table \ref{tab:best_model_param}, the $T_{\rm eff}$ ranges of the best-fitting SM08 (1000--1200 K) and BT-Settl (1000--1150 K) models are now more consistent with each other. The SM08 models still indicate a somewhat higher surface gravity ($\log g=4.0$--5.0) than BT-Settl ($\log g=3.5$--4.0). The best-fitting SM08 models all have sedimentation efficiencies of $f_{\rm sed}=3$.

We further assess the quality of the photospheric fits by computing synthetic fluxes in each photometric band. This allows us to also gauge the models in wavelength regions where we do not have spectroscopic coverage.  We compare the flux offsets between the observed photometry (shown in various colors in Figure~\ref{fig:SED}) and the synthetic photometry (shown in pink in Figure~\ref{fig:SED}).  We calculate the synthetic photometry $F_{\rm syn}$ over a band pass ($\lambda_A, \lambda_B$) by convolving the filter response $S_{\lambda}$ with the synthetic spectrum $F_{\lambda}$:
\begin{equation}
F_{\rm syn} = \dfrac{\int_{\lambda_A}^{\lambda_B}F_{\lambda}S_{\lambda}d\lambda}{\int_{\lambda_A}^{\lambda_B}S_{\lambda}d\lambda}
\label{eq:synt_phot}
\end{equation}
The synthetic photometry in Figure \ref{fig:SED} corresponds to the median value for the five best-fit models (SM08 in left panel, BT-Settl in right panel). The vertical error bars on the synthetic photometry correspond to the 86\% central confidence interval.

It is evident that, when seeking to reproduce the full SED, both sets of models face challenges in three distinct areas: (1) the red optical continuum (for SM08 models), (2) the predicted strengths of CH$_4$ absorption (notably the 3.3~$\mu$m and 8.0~$\mu$m bands for the SM08 models; the 1.6~$\mu$m band for BT-Settl), and (3) the CO absorption band head at 4.6~$\mu$m.  The red optical continuum poses known difficulties because of problems with predicting the strength of the highly pressure-broadened optical \ion{K}{1} and \ion{Na}{1} lines \citep{Burrows-Volobuyev2003,Cushing_etal2008,Allard_etal2016}. The SM08 models over-predict CH$_4$ absorption strengths and under-predict the CO strength, which is explained by the assumption of chemical equilibrium in these models. BT-Settl models, which incorporate vertical mixing to produce non-equilibrium chemistry, fit overall better the observed CH$_4$ absorption, although they under-predict its strength at 1.6~$\mu$m and slightly over-predict the strength of the CO absorption. 

Surprisingly, the clear-atmosphere Sonora 2018 and ATMO 2020 models with chemical equilibrium reproduce the CO strength quite well. However, they significantly over-predict the CH$_4$ absorption at 3.3~$\mu$m and 8.0~$\mu$m (Appendix \ref{sec:cloudless_model_fits}, Fig.~\ref{fig:SED_test}).

The challenges faced by the models in reproducing the depths of the strongest CH$_4$ (at 3.3~$\mu$m) and CO (at 4.6~$\mu$m) absorption bands point to the importance of non-equilibrium chemistry in the atmosphere of HN~Peg~B. The problem is not unique to HN~Peg~B; it is pertinent to early-T dwarfs in general \citep[e.g.,][]{Phillips_etal2020}, as the 3--6~$\mu$m colors and SED of HN~Peg~B are very similar to those of other typical early-T dwarfs \citep{Luhman_etal2007}. 

\begin{table*}
\caption{Best-fitting model spectra to the near- and mid-IR spectrophotometry of HN~Peg~B.}
  \small
  \label{tab:best-fitting_model_spectra}
  \begin{threeparttable}
 	\begin{tabular}{@{\extracolsep{2pt}}llcccclccc@{}}
    \toprule
	Observed                    &   \multicolumn{5}{c}{SM08 Model}                           & \multicolumn{4}{c}{BT-Settl Model} \\
	\cline{2-6}
	\cline{7-10}
	Spectrum                    & Spectrum         & $\chi^2_r$ & $T_{\rm eff}$ (K) & $\log g$ & $f_{\rm sed}$ & Spectrum     & $\chi^2_r$ & $T_{\rm eff}$ (K)  & $\log g$ \\
    \midrule
    \multirow{5}{*} {NIRSPEC}   & sp\_t1100g100f4   & 1.1      & 1100      & 4.0      & 4         & lte010.5-3.0 & 2.0       & 1050       & 3.0      \\
                                & sp\_t1200g100f4   & 1.2      & 1200      & 4.0      & 4         & lte011-3.0   & 2.1       & 1100       & 3.0      \\
                                & sp\_t1000g300f3   & 1.4      & 1000      & 4.5      & 3         & lte011-3.5   & 2.2       & 1100       & 3.5      \\
                                & sp\_t1100g100f3   & 1.4      & 1100      & 4.0      & 3         & lte010-2.5   & 2.2       & 1000       & 2.5      \\
                                & sp\_t1100g300f3   & 1.5      & 1100      & 4.5      & 3         & lte011.5-3.5 & 2.4       & 1150       & 3.5      \\
    \midrule
    \multirow{5}{*} {SpeX}      & sp\_t1300g100f4   & 3.8      & 1300      & 4.0      & 4         & lte012-5.0   & 3.4       & 1200       & 5.0      \\
                                & sp\_t1400g300f4   & 4.4      & 1400      & 4.5      & 4         & lte013-5.5   & 5.1       & 1300       & 5.5      \\
                                & sp\_t1300g300f4   & 5.2      & 1300      & 4.5      & 4         & lte011-4.0   & 6.3       & 1100       & 4.0      \\
                                & sp\_t1400g1000f4  & 5.6      & 1400      & 5.0      & 4         & lte011.5-5.0 & 7.8       & 1150       & 5.0      \\
                                & sp\_t1200g100f4   & 5.7      & 1200      & 4.0      & 4         & lte010-3.5   & 9.5       & 1000       & 3.5      \\
    \midrule
    \multirow{5}{*} {NIRI}      & sp\_t1300g3000f2  & 2.6      & 1300      & 5.5      & 2         & lte014-5.5   & 2.4       & 1400       & 5.5      \\
                                & sp\_t1300g1000f2  & 2.7      & 1300      & 5.0      & 2         & lte012-4.0   & 3.0       & 1200       & 4.0      \\
                                & sp\_t1400g3000f3  & 2.7      & 1400      & 5.5      & 3         & lte012-4.5   & 3.2       & 1200       & 4.5      \\
                                & sp\_t900g3000f1   & 2.8      & 900       & 5.5      & 1         & lte011.5-4.0 & 3.5       & 1150       & 4.0      \\
                                & sp\_t1500g300f4   & 2.8      & 1500      & 4.5      & 4         & lte011.5-3.5 & 3.8       & 1150       & 3.5      \\
    \midrule                            
    \multirow{5}{*} {IRS}       & sp\_t900g3000f1   &  6.5     & 900       & 5.5      & 1         & lte011-4.0   & 4.9       & 1100       & 4.0      \\
                                & sp\_t900g300f1    &  8.1     & 900       & 4.5      & 1         & lte010.5-3.5 & 5.2       & 1050       & 3.5      \\
                                & sp\_t1100g3000f2  & 10.3     & 1100      & 5.5      & 2         & lte012-5.5   & 5.2       & 1200       & 5.5      \\
                                & sp\_t1200g3000f3  & 10.4     & 1200      & 5.5      & 3         & lte011.5-5.0 & 5.3       & 1150       & 5.0      \\
                                & sp\_t900g100f1    & 10.7     & 900       & 4.0      & 1         & lte011-3.5   & 5.4       & 1100       & 3.5      \\
    \midrule                            
    \multirow{5}{*} {Full SED$^a$} & sp\_t1100g100f3 & 11.3     & 1100      & 4.0      & 3         & lte010.5-3.5 &  7.3      & 1050       & 3.5      \\
                                & sp\_t1200g100f3   & 12.5     & 1200      & 4.0      & 3         & lte010-3.5   &  7.3      & 1000       & 3.5      \\
                                & sp\_t1100g300f3   & 14.1     & 1100      & 4.5      & 3         & lte011-4.0   &  7.7      & 1100       & 4.0      \\
                                & sp\_t1100g1000f3  & 14.4     & 1100      & 5.0      & 3         & lte011.5-4.0 & 12.8      & 1150       & 4.0      \\
                                & sp\_t1000g1000f3  & 15.8     & 1000      & 5.0      & 3         & lte011-3.5   & 13.2      & 1100       & 3.5      \\
    \bottomrule
 	\end{tabular}
  \begin{tablenotes}[para,flushleft]
	\hspace{20ex}$^a$Including all spectra and photometry of HN~Peg~B.
  \end{tablenotes}
 \end{threeparttable}
\end{table*}

\begin{table*}
\begin{adjustwidth}{-0.5in}{-1in}
\caption{Parameters of the best-fitting model spectra to the spectrophotometry of HN~Peg~B.}
  \small
  \label{tab:best_model_param}
  \begin{threeparttable}
 	\begin{tabular}{@{\extracolsep{2pt}}lcccccclcc@{}}
    \toprule
	Observed  & Wavelength   & Resolution     & \multicolumn{4}{c}{SM08 Model}                    & \multicolumn{3}{c}{BT-Settl Model} \\
	\cline{4-7}
	\cline{8-10}
	Spectrum     & Range        &                & $\chi^2_r$  & $T_{\rm eff}$  & $\log g$  & $f_{\rm sed}$  & $\chi^2_r$ & $T_{\rm eff}$  & $\log g$ \\
                 & ($\mu$m)     &                &             &   (K)      &           &            &            &     (K)    &          \\
    \midrule
    NIRSPEC      & 1.143--1.375 & $\approx2300$  & 1.1--1.5    & 1000--1200 & 4.0--4.5  & 3--4       &  2.0--2.4  & 1000--1150 & 2.5--3.5 \\
    SpeX         & 0.8--2.5     & $\approx100$   & 3.8--5.7    & 1200--1400 & 4.0--5.0  & 4          &  3.4--9.5  & 1000--1300 & 3.5--5.5 \\
    NIRI         & 2.96--4.07   & $\approx400$   & 2.6--2.8    & 900--1500  & 4.5--5.5  & 1--4       &  2.4--3.8  & 1150--1400 & 3.5--5.5 \\
    IRS          & 5.2--14.2    & $\sim60$--130  & 6.5--10.7   & 900--1200  & 4.0--5.5  & 1--3       &  4.9--5.4  & 1050--1200 & 3.5--5.5 \\
	Full SED$^a$ & 0.8--26.3    & \nodata        & 11.3--15.8  & 1000--1200 & 4.0--5.0  & 3          &  7.3--13.2 & 1000--1150 & 3.5--4.0 \\
    \bottomrule
 	\end{tabular}
  \begin{tablenotes}[para,flushleft]
	\hspace{20ex}$^a$Including all spectra and photometry of HN~Peg~B.
  \end{tablenotes}
  \end{threeparttable}
\end{adjustwidth}
\end{table*}

\subsubsection{Comparisons of the Best-fitting Model Spectra: Parameter Estimates from Spectrophotometric Fitting}
\label{sec:comparison_best-fitting_models}

The parameters of the five best-fitting photospheres from the SM08 and BT-Settl models to each of the data sets presented in Sections~\ref{sec:NIRSPEC_model_comparisons}--\ref{sec:all_model_comparisons} are listed in Table~\ref{tab:best-fitting_model_spectra}.  We summarize the ranges of these parameters for each data set in Table~\ref{tab:best_model_param}.

The best fits, in terms of the lowest $\chi^2_r$, are obtained for the Keck/NIRSPEC and Gemini/NIRI spectra.  However, rather than using $\chi^2_r$ as guidance on which data set might produce the most reliable physical parameters, we adopt the results from the best fits to the full SED (all spectra and photometry).  These fits take advantage of the best possible overall flux constraint, and so deliver the most accurate estimates for $T_{\rm eff}$ and $\log g$. These are consistent with the estimates from all individual spectral ranges (Table~\ref{tab:best_model_param}). At the same time, fitting the entire data set also produces the highest $\chi^2_r$ values, as can be expected.  As summarized in Table~\ref{tab:HN_PegB_parameters}, we find: $T_{\rm eff}$=1000--1200 K, $\log g$=4.0--5.0 and $f_{\rm sed}$=3 (SM08) or $T_{\rm eff}$=1000--1150 K and $\log g$=3.5--4.0 (BT-Settl).

\begin{figure*}
	\centering
	\begin{tabular}{cccccccc}
		\subfloat{\includegraphics[width=.40\linewidth]{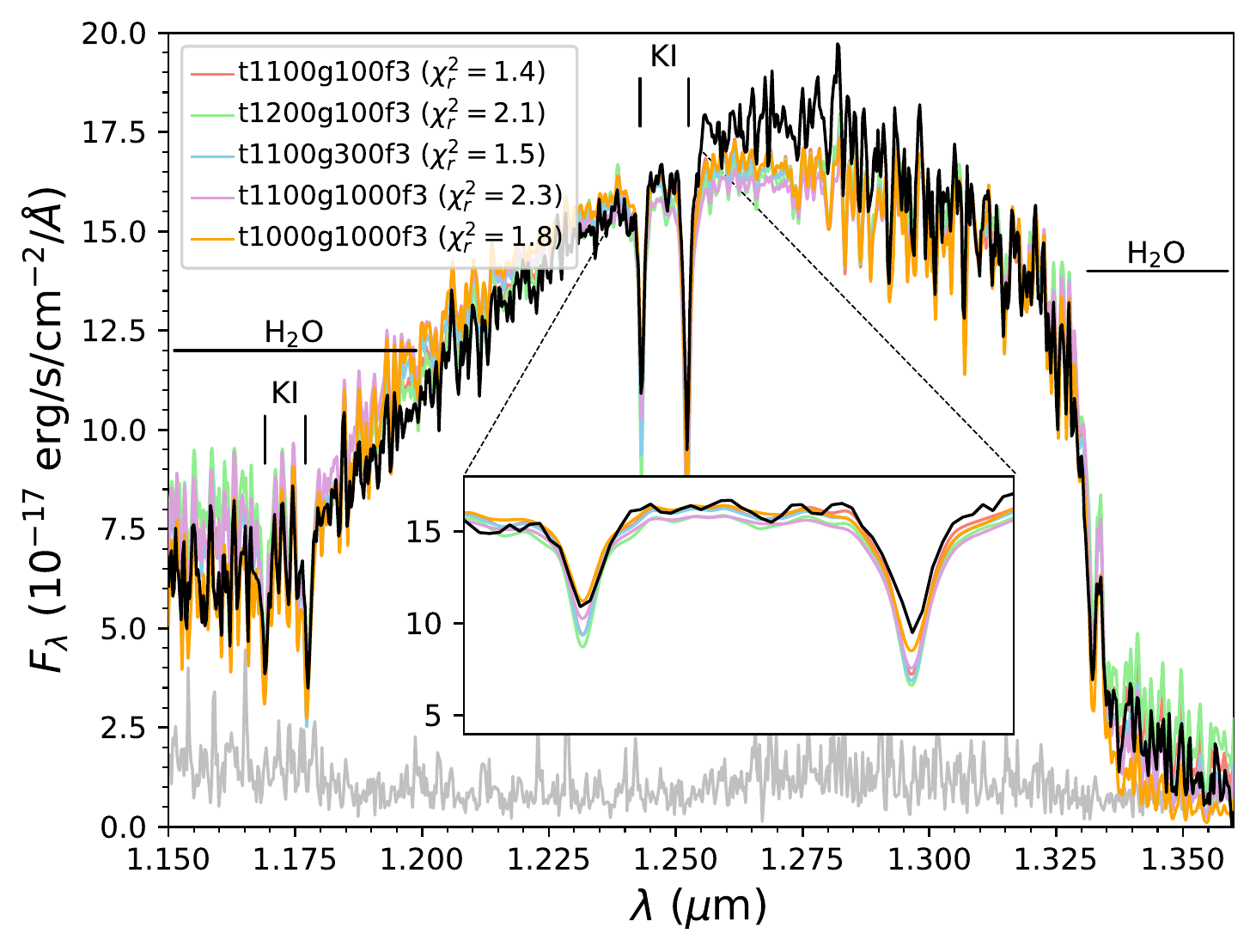}} &
		\subfloat{\includegraphics[width=.40\linewidth]{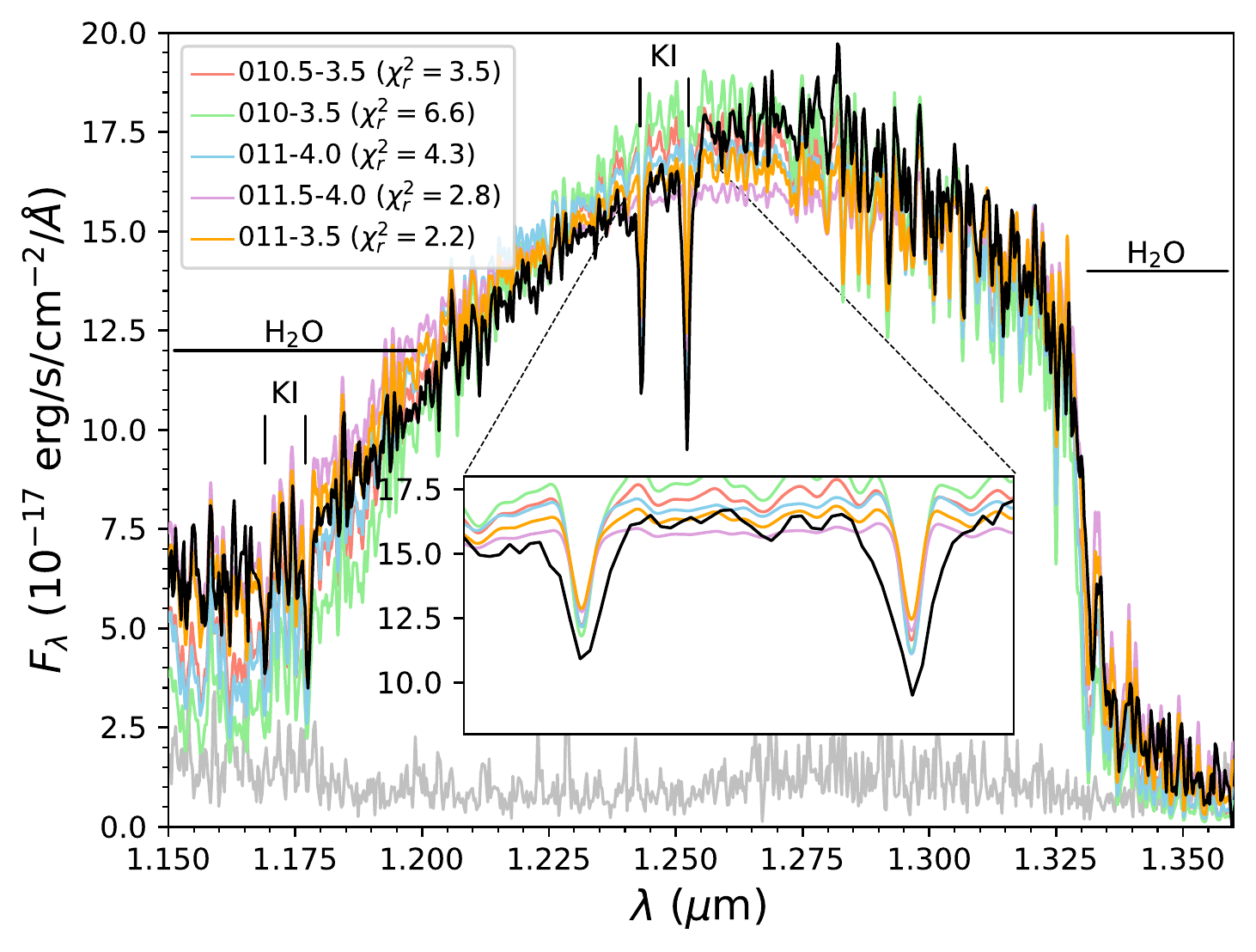}} & \\
		\subfloat{\includegraphics[width=.40\linewidth]{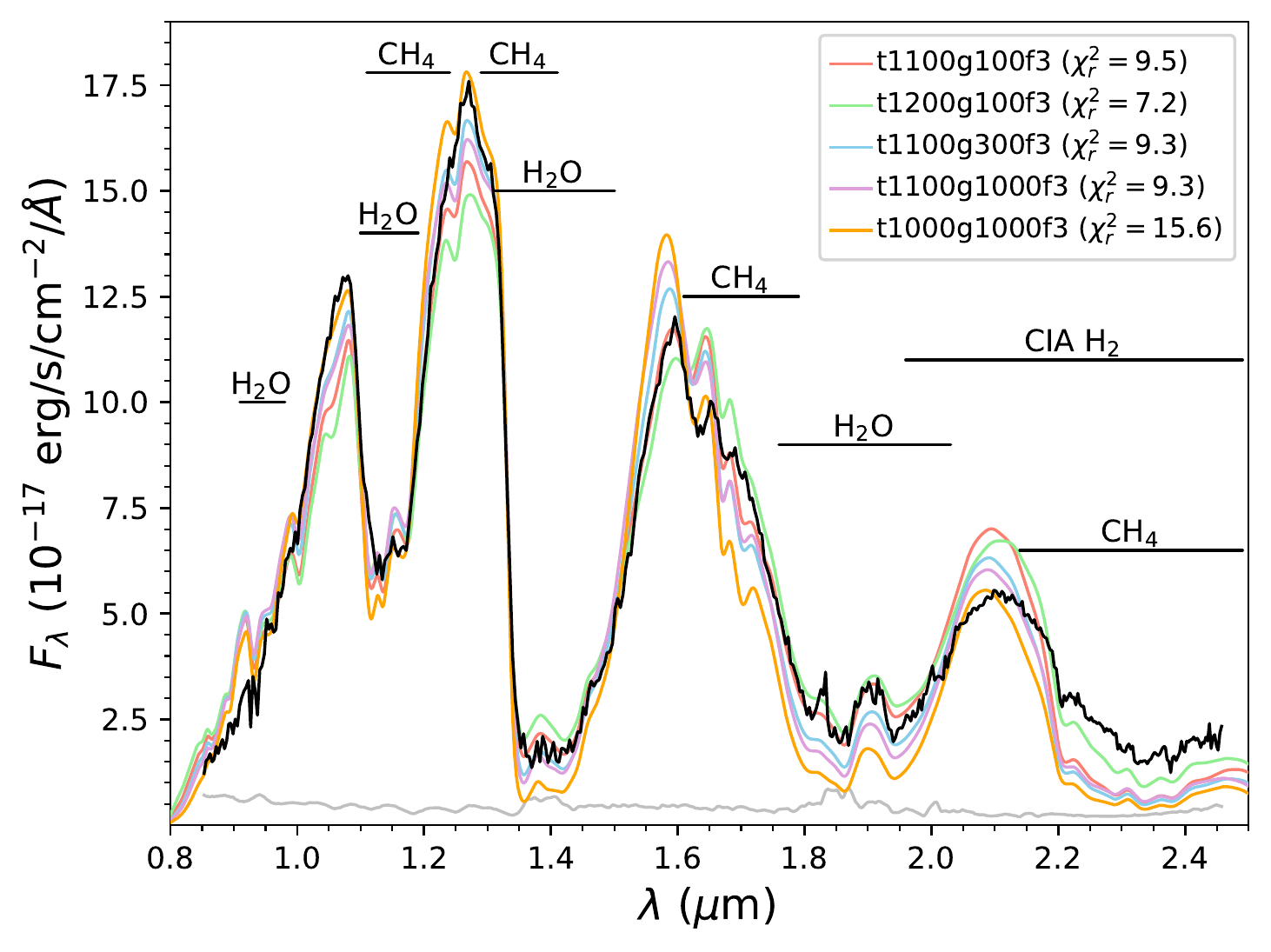}} &
		\subfloat{\includegraphics[width=.40\linewidth]{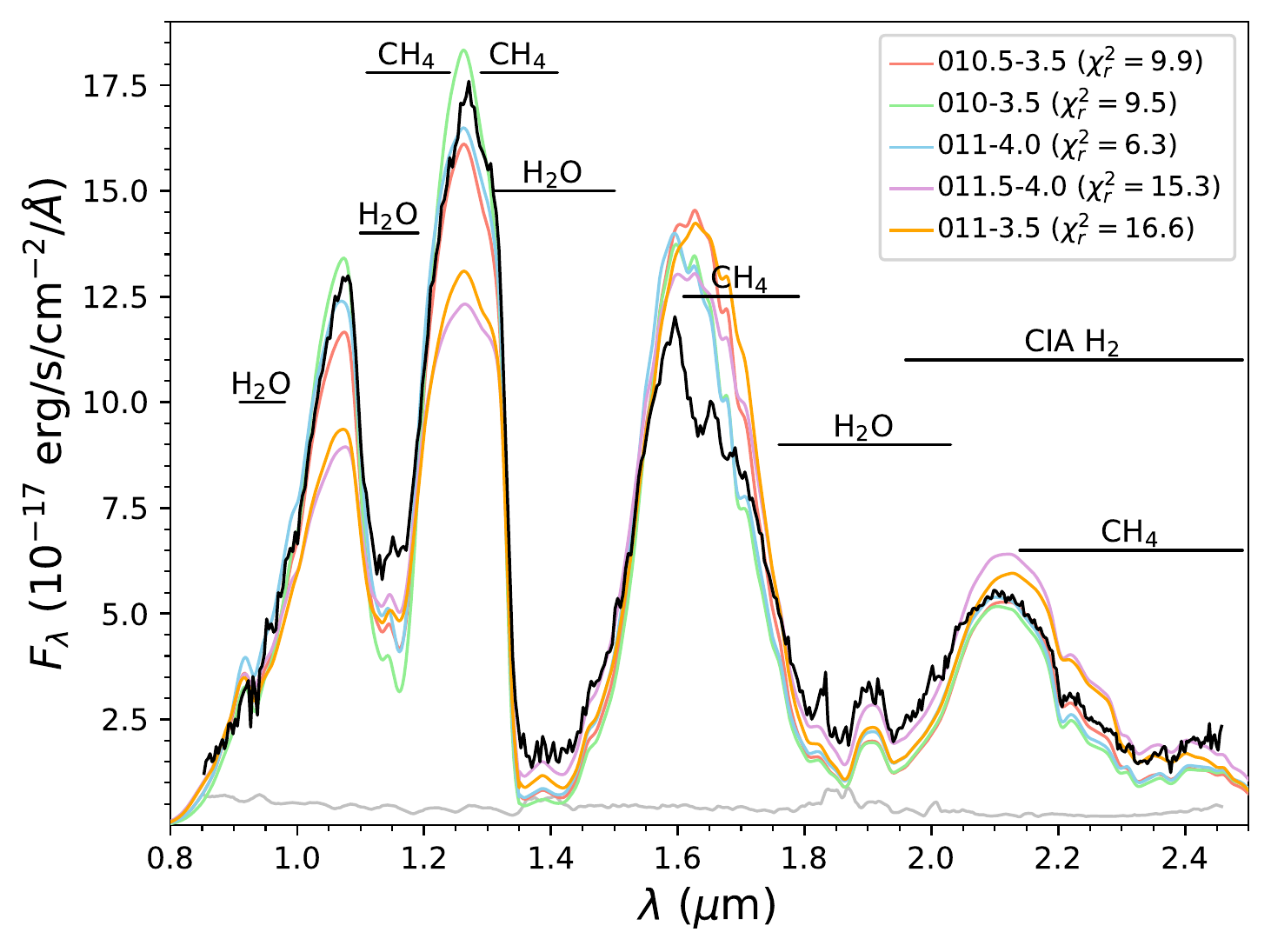}} & \\
		\subfloat{\includegraphics[width=.40\linewidth]{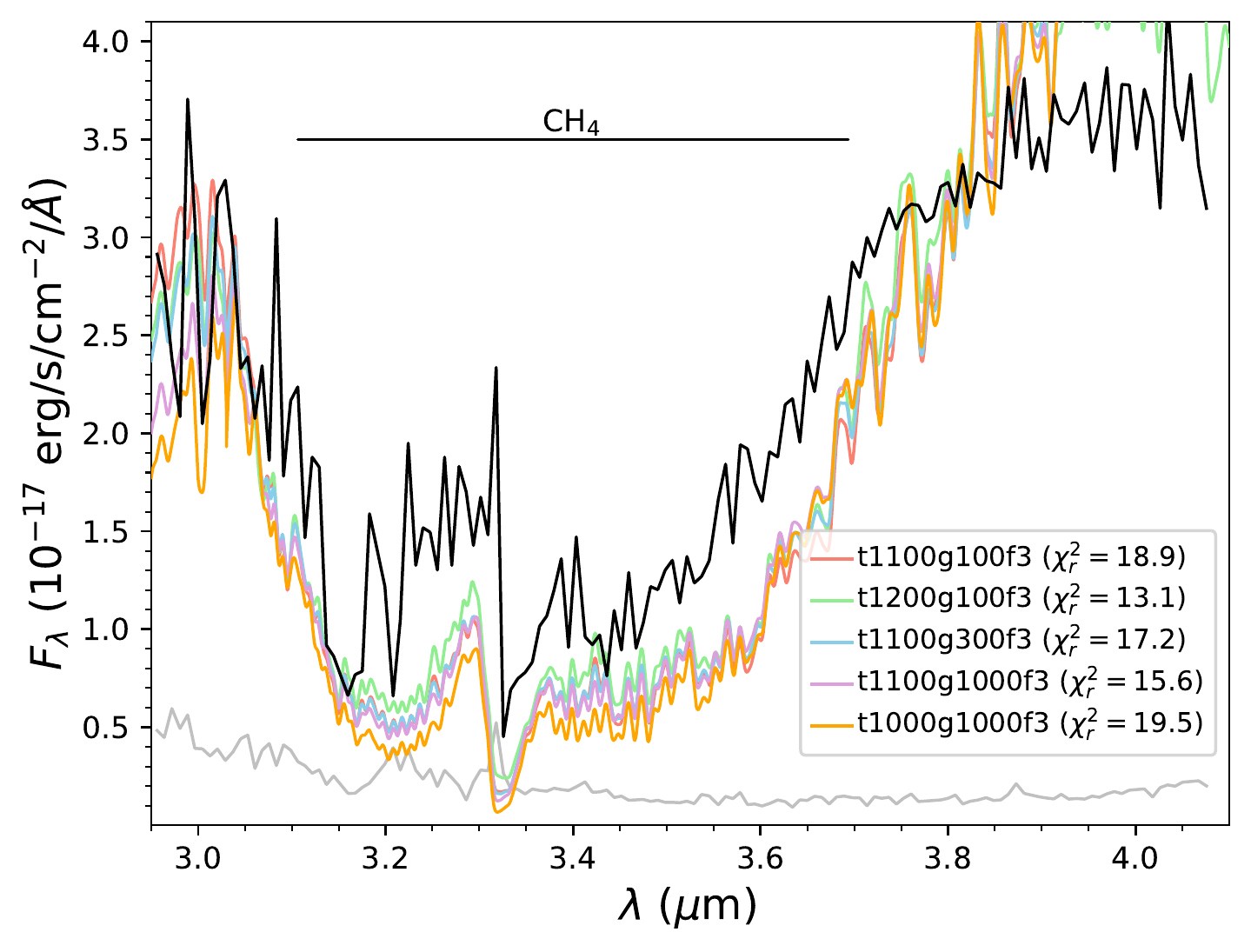}} &
		\subfloat{\includegraphics[width=.40\linewidth]{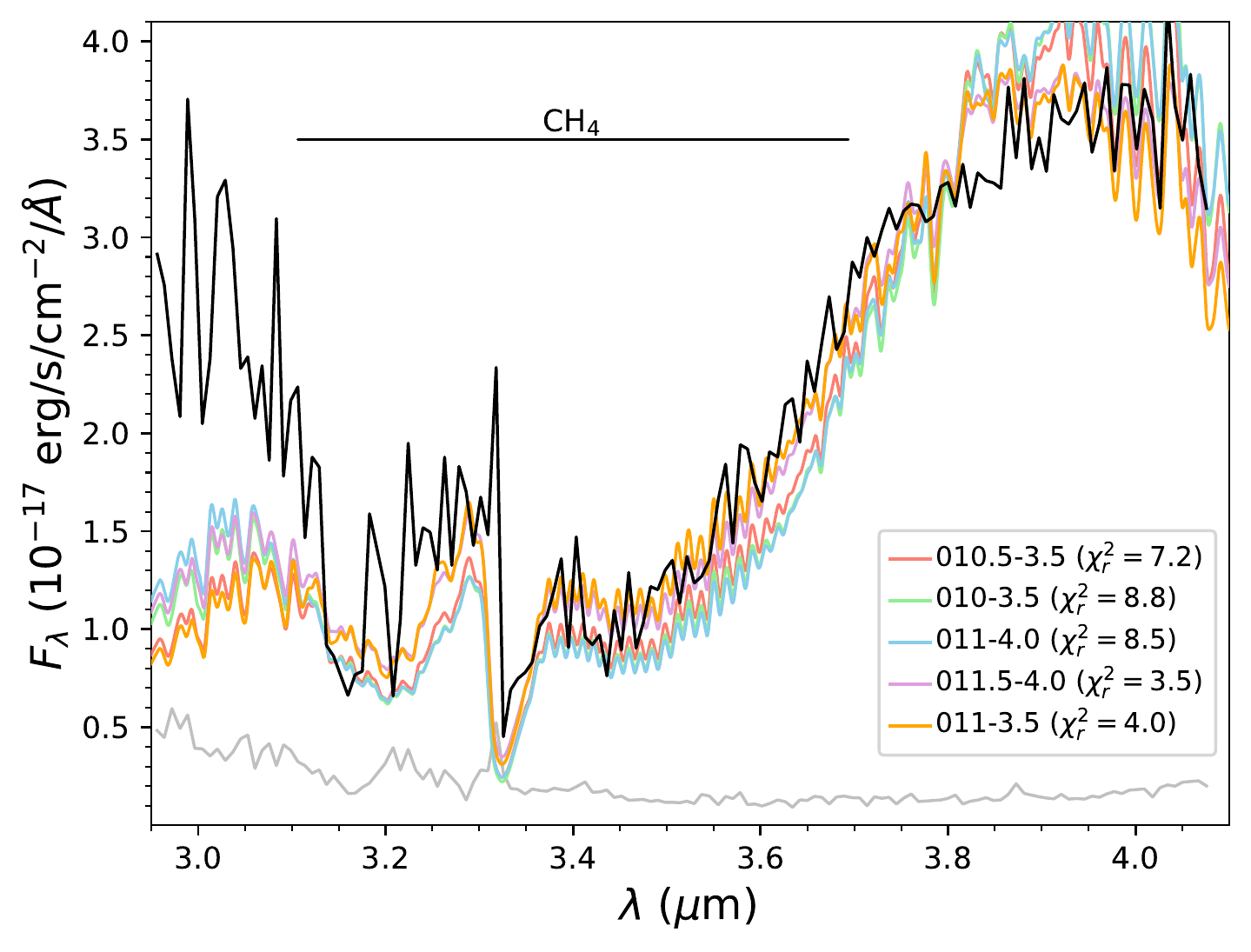}} \\
		\subfloat{\includegraphics[width=.40\linewidth]{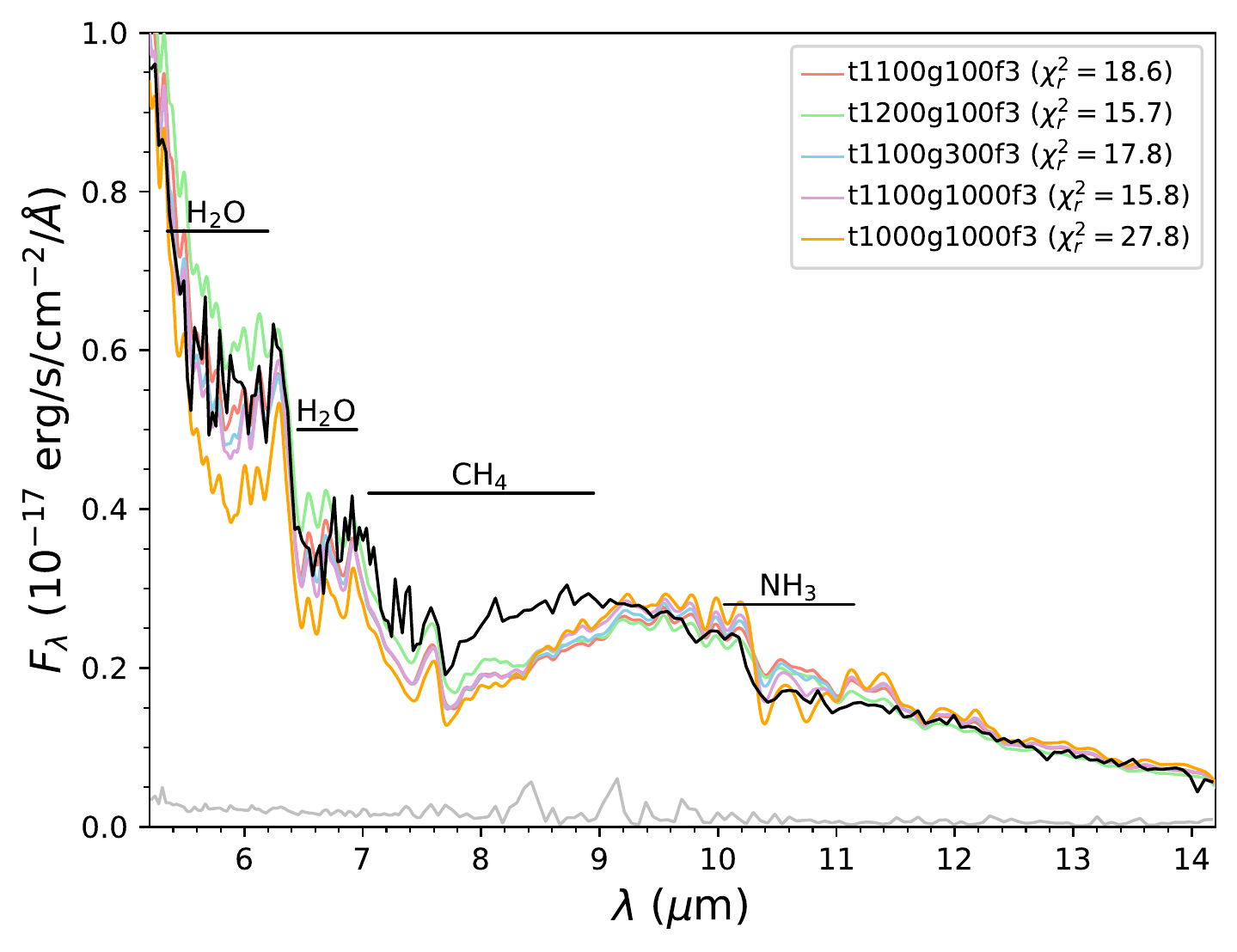}} &
		\subfloat{\includegraphics[width=.40\linewidth]{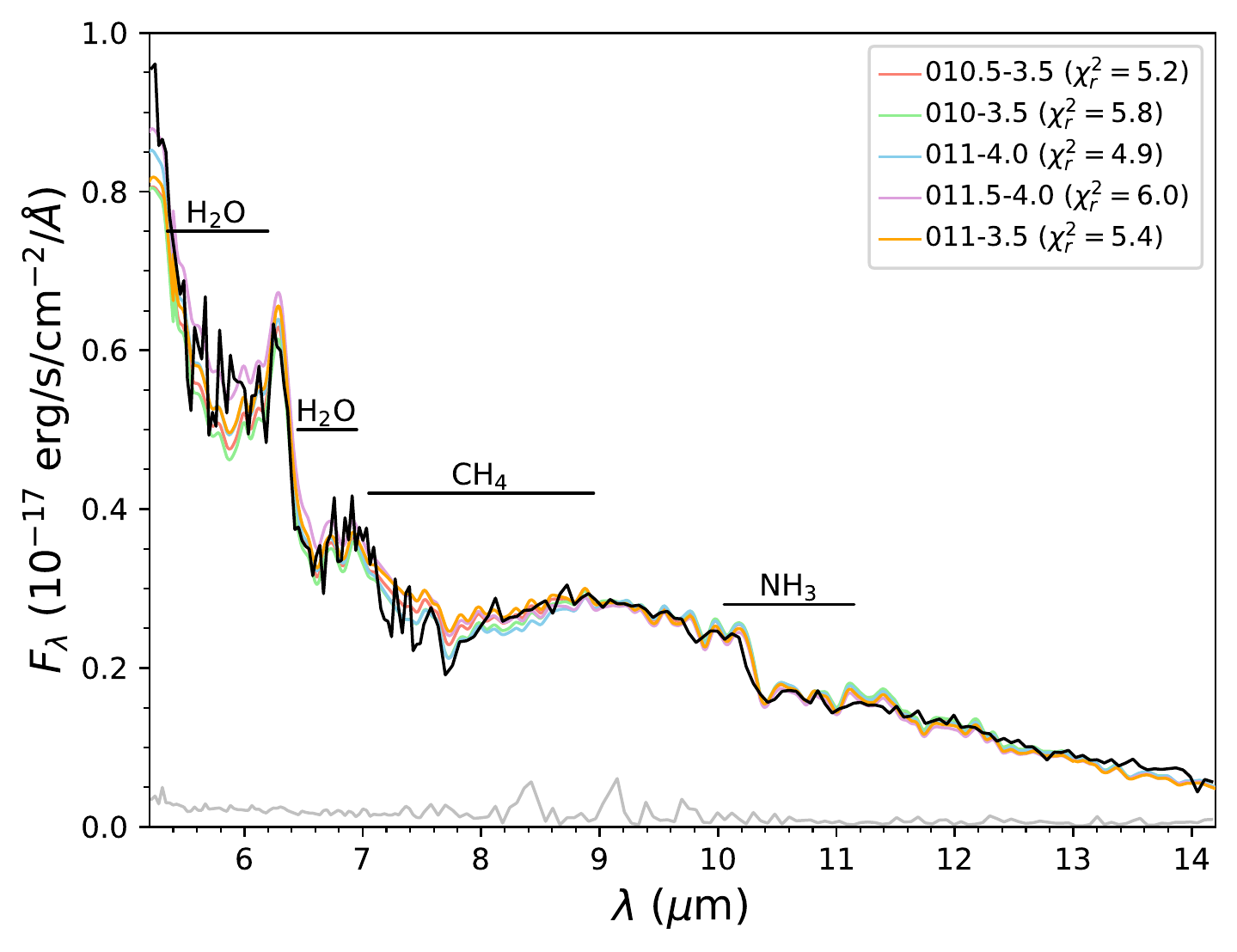}}
	\end{tabular}
	\caption{Overall best-fitting SM08 (left column) and BT-Settl (right column) model photospheres compared to the individual NIRSPEC (first row), SpeX (second row), NIRI (third row) and IRS (last row) spectra.}
	\label{fig:comparison_best-fitting_models}
\end{figure*}

In Figure \ref{fig:comparison_best-fitting_models} we show how the overall best-fitting SM08 and BT-Settl model photospheres match each of the observed spectra separately. The models reproduce the individual spectra adequately, and in most cases the $\chi^2_r$ is within a factor of two from the $\chi^2_r$ for the best-fitting models to the individual spectra.  A notable exception is the overall best-fitting SM08 models when compared to the NIRI spectrum.  In this case the $\chi^2_r$ is greater by a factor of six compared to the SM08 fits to the NIRI spectrum alone (Figure~\ref{fig:model_comparisons_NIRI_all_best}; Table~\ref{tab:best-fitting_model_spectra}).  The reason is the previously mentioned (Section~\ref{sec:all_model_comparisons}) over-prediction of the 3.3~$\mu$m methane absorption strength by the SM08 models.  A similar discrepancy, also caused by over-predicted methane absorption, is seen in the way the overall best-fitting SM08 model photospheres reproduce the {\it Spitzer}/IRS spectrum (bottom left panel of Figure~\ref{fig:comparison_best-fitting_models}).

\subsection{Semi-empirical and Evolutionary Model-based Physical Parameter Estimates for HN~Peg~B}
\label{sec:phy_par}
Herein we calculate the bolometric luminosity of HN~Peg~B, we obtain semi-empirical estimates for the radius and effective temperature, and use evolutionary models to estimate the surface gravity, mass, and age.

Our empirical SED includes 98\% of the HN~Peg~B bolometric luminosity (Figure~\ref{fig:SED}), and the \textit{Gaia} distance to HN~Peg~A \citep[$18.133\pm0.012$~pc;][]{GaiaCollaboration2021} is unprecedentedly accurate and precise.  We can thus obtain a highly reliable estimate of the bolometric luminosity $L_{\rm bol}$ of HN~Peg~B. For the remaining 2\% of the flux contained in the $\lambda<0.85\ \mu$m, $2.5< \lambda\ (\mu\rm{m}) <3.0$ and $\lambda>26.3\ \mu$m regions, we integrate the best-fitting SM08 and BT-Settl model photospheres.  The derived luminosity is $\log(L_{\rm bol}/L_\odot)=-4.79\pm0.02$, regardless of which model family is used to complement the wavelength coverage.\footnote{We use $L_\odot=3.844\times10^{33}$ erg s$^{-1}$ \citep{Bahcall_etal1995}} The uncertainty is obtained by taking into account the measured flux density uncertainties, the distance error, the dispersion (68\% central confidence interval) of the best-fitting model photospheres, and the known 0.8\%--1.2\% variability of HN~Peg~B \citep{Metchev_etal2015,Zhou_etal2018}. Because the various SED data were obtained at different epochs, the overall effect on the bolometric uncertainty due to the variability is negligible. We verified this by performing a Monte Carlo simulation that randomly scaled the spectrophotometry within the variability amplitudes at different wavelengths and by measuring the resulting scatter in $10^4$ $L_{\rm bol}$ estimates. The resulting bolometric luminosity is in agreement with earlier estimates based on sparser wavelength coverage: $-4.77\pm0.03$ \citep{Luhman_etal2007}, $-4.76\pm0.02$ \citep{Leggett_etal2008}, and $-4.82\pm0.01$ \citep{Filippazzo_etal2015}. 

Since the dilution factor $\alpha$ in the $\chi^2_r$ definition in Equation \ref{eq:chi2_r} is $\alpha=(R/d)^2$, we also compute the radius of HN~Peg~B, given the known distance to the system and the normalization constants determined from the best-fitting model photospheres. We find $R=1.06_{-0.06}^{+0.08}\ R_{\rm Jup}$ ($0.107_{-0.006}^{+0.008}\ R_\odot$) based on the SM08 models or $R=1.13_{-0.06}^{+0.09}\ R_{\rm Jup}$ ($0.114_{-0.006}^{+0.009}\ R_\odot$) based on the BT-Settl models. The values correspond to the medians of the estimates from the five best photospheric fits from each model family, and the uncertainties correspond to the 68\% confidence interval. The variability of HN~Peg~B contributes only 0.1\% to the radius uncertainty budget. Consequently, from the Stefan-Boltzmann (S-B) Law, we find $T_{\rm eff}=1126_{-36}^{+35}$ K (SM08) or $T_{\rm eff}=1090_{-41}^{+32}$ K (BT-Settl).

The above estimates for $R$ and $T_{\rm eff}$ are semi-empirical, as we have relied on best-fit model photospheres to determine the dilution factor $\alpha$, and hence the radius $R$.  We now use $L_{\rm bol}$ and $R$ with evolutionary models to interpolate the surface gravity, mass, and age.  We use the SM08 and Sonora 2018 \citep[][Marley et al.\ 2020, in preparation]{Marley_etal2018} evolutionary models together with the SM08-based estimate for the radius, and the COND03 \citep{Baraffe_etal2003} and ATMO 2020 \citep{Phillips_etal2020} evolutionary models with the BT-Settl-based estimate for the radius.  In this manner, the model photospheres and the evolutionary models derive from the same family of models (Ames/Los Alamos vs.\ Lyon/Exeter group), each with a self-consistent set of opacity assumptions. The evolutionary model comparisons also deliver estimates for $T_{\rm eff}$ that are, by design, very similar to the S-B Law determinations above.  

The results for the fundamental parameters of HN~Peg~B from the evolutionary model interpolations are shown in Table \ref{tab:HN_PegB_parameters}.  The parameter uncertainties were obtained from Monte Carlo simulations to sample the distributions of $L_{\rm bol}$ and $R$.  For $L_{\rm bol}$ we used $10^4$ values randomly sampled from a normal distribution defined by the mean $L_{\rm bol}$ and its uncertainty.  For $R$ we generated $10^4$ values by random sampling from the parameter's cumulative distribution.  We used these simulated $L_{\rm bol}$ and $R$ values with the evolutionary models to interpolate $T_{\rm eff}$, $\log g$, mass, and age, and considered the 68\% central confidence interval (i.e., comparable to $\pm$1$\sigma$ for Gaussian errors) to obtain the uncertainty of the derived parameters. 

Table \ref{tab:HN_PegB_parameters} reveals that the two model families produce very similar results, although there are some consistent differences. The Ames/Los Alamos models yields a 6\% smaller radius, 3\% higher $T_{\rm eff}$, 70\% higher surface gravity ($\log g=4.77$ vs.\ 4.54), 50\% higher mass, and a factor of $\sim$2 older age than the Lyon/Exeter models.  The systematic differences are comparable to the parameter uncertainties.  As our adopted parameters of HN~Peg~B we take the average of the two model families. We consider the half-difference between the two sets of mean values as a systematic uncertainty and the averages of the upper and lower uncertainties as the final random uncertainties on each estimate. The total uncertainties for the adopted parameters correspond to the quadrature sum of the systematic and the final random uncertainties.  For $T_{\rm eff}$ we consider the results from the S-B Law as most accurate, and use only them, rather than the (very similar) results from the evolutionary model fits.  The bottom section of Table \ref{tab:HN_PegB_parameters} lists our adopted parameter values for HN~Peg~B and the systematic and random uncertainties.

\begin{table*}
\caption{HN~Peg~B Physical Parameters Determined in this Study.}
  \label{tab:HN_PegB_parameters}
  \begin{threeparttable}
	\begin{tabularx}{\linewidth}{lccccccc}
    \toprule
	Origin                               & $\log (L_{\rm bol}/L_\odot$) & $R$                    & $T_{\rm eff}$      & $\log g$               & $f_{\rm sed}$ & mass                     & $\log(\textrm{age})$   \\
	                                     &                              & ($R_{\rm Jup}$)        & (K)                &                        &               & ($M_{\rm Jup}$)          &                        \\ 
    \midrule
	SM08 Model Spectra Fits$^a$          & \nodata                & $1.06_{-0.06}^{+0.08}$    & 1000--1200         & 4.0--5.0               & 3         & \nodata                   & \nodata                \\
	BT-Settl Model Spectra Fits$^a$      & \nodata                & $1.13_{-0.06}^{+0.09}$    & 1000--1150         & 3.5--4.0               & \nodata   & \nodata                   & \nodata                \\
	Spectral Energy Distribution         & $-4.79\pm0.02$         & \nodata                   & \nodata            & \nodata                & \nodata   & \nodata                   & \nodata                \\
	S-B Law with $R$ from SM08 model     & \nodata                & \nodata                   & $1126_{-36}^{+35}$ & \nodata                & \nodata   & \nodata                   & \nodata                \\
	S-B Law with $R$ from BT-Settl model & \nodata                & \nodata                   & $1090_{-41}^{+32}$ & \nodata                & \nodata   & \nodata                   & \nodata                \\
	SM08 Evolutionary Model$^b$          & \nodata                & \nodata                   & $1128_{-39}^{+32}$ & $4.78_{-0.20}^{+0.14}$ & \nodata   & $26_{-7}^{+6}$            & $8.76_{-0.29}^{+0.22}$ \\
	Sonora 2018 Evolutionary Model$^b$   & \nodata                & \nodata                   & $1124_{-37}^{+34}$ & $4.76_{-0.20}^{+0.15}$ & \nodata   & $25_{-6}^{+6}$            & $8.59_{-0.22}^{+0.22}$ \\
	COND03 Evolutionary Model$^b$        & \nodata                & \nodata                   & $1096_{-41}^{+34}$ & $4.54_{-0.25}^{+0.16}$ & \nodata   & $18_{-6}^{+6}$            & $8.29_{-0.37}^{+0.24}$ \\
	ATMO 2020 Evolutionary Model$^b$     & \nodata                & \nodata                   & $1090_{-41}^{+34}$ & $4.54_{-0.25}^{+0.17}$ & \nodata   & $17_{-5}^{+5}$            & $8.27_{-0.47}^{+0.19}$ \\
    \midrule
	Adopted Parameters$^c$               & $-4.79\pm0.02$         & $1.10_{-0.07}^{+0.09}$    & $1108_{-43}^{+37}$ & $4.66_{-0.25}^{+0.20}$ & 3         & $21_{-7}^{+7}$            & $8.48_{-0.39}^{+0.29,f}$ \\
	Systematic Uncertainties$^d$         & \nodata                & $\pm$0.04                 & $\pm$18            & $\pm$0.11              & \nodata   & $\pm$4                    & $\pm$0.20              \\
	Random Uncertainties$^e$             & \nodata                & $_{-0.06}^{+0.08}$        & $_{-38}^{+33}$     & $_{-0.22}^{+0.16}$     & \nodata   & $_{-6}^{+6}$              & $_{-0.34}^{+0.22}$     \\
    \bottomrule
 	\end{tabularx}
	\begin{tablenotes}[para,flushleft]
	$^a$ Obtained by fitting model photospheres to the full SED and minimizing $\chi^2_r$. \\
	$^b$ Obtained by interpolating $L_{\rm bol}$ and $R$ into the respective evolutionary models. \\
	$^c$ For $R$: the mean of the estimates from the SM08 and BT-Settl model photospheres. For $T_{\rm eff}$: the mean of the two S-B Law determinations. For $\log g$, mass, and age: the mean of the four evolutionary-model estimates. The uncertainties correspond to the quadrature sum of systematic and random uncertainties.\\
	$^d$ Obtained as the half-difference between the mean values of the two model families.\\
	$^e$ Obtained as averages of the lower or upper random uncertainties of the parameters (see Section~\ref{sec:phy_par} for details). \\
	$^f$ The adopted age corresponds to $300_{-170}^{+280}$ Myr.
	\end{tablenotes}
  \end{threeparttable}
\end{table*}

\section{DISCUSSION}
\label{sec:discussion}
We have already discussed two of the principal findings in our study.  In Section~\ref{sec:results:ki} we analyzed the sensitivity of the $J$-band \ion{K}{1} doublet to surface gravity.  We concluded that in T2--T3 dwarfs with moderate to high gravities the \ion{K}{1} line strengths are insensitive to surface gravity. In Section~\ref{sec:all_model_comparisons} we compared the model photosphere predictions to the full 0.8--26.3 $\mu$m spectrophotometry of HN~Peg~B, and found that the models face challenges in predicting the extent of the CH$_4$/CO chemical disequilibrium in this young early-T dwarf, although BT-Settl model photospheres perform better on average.  Below we discuss additional findings on the fundamental parameters of HN~Peg~B from Section~\ref{sec:phy_par}.

\subsection{The Luminosity, Effective Temperature, and Radius of HN Peg B}
\begin{figure*}
	\centering
	\includegraphics[width=0.6\linewidth]{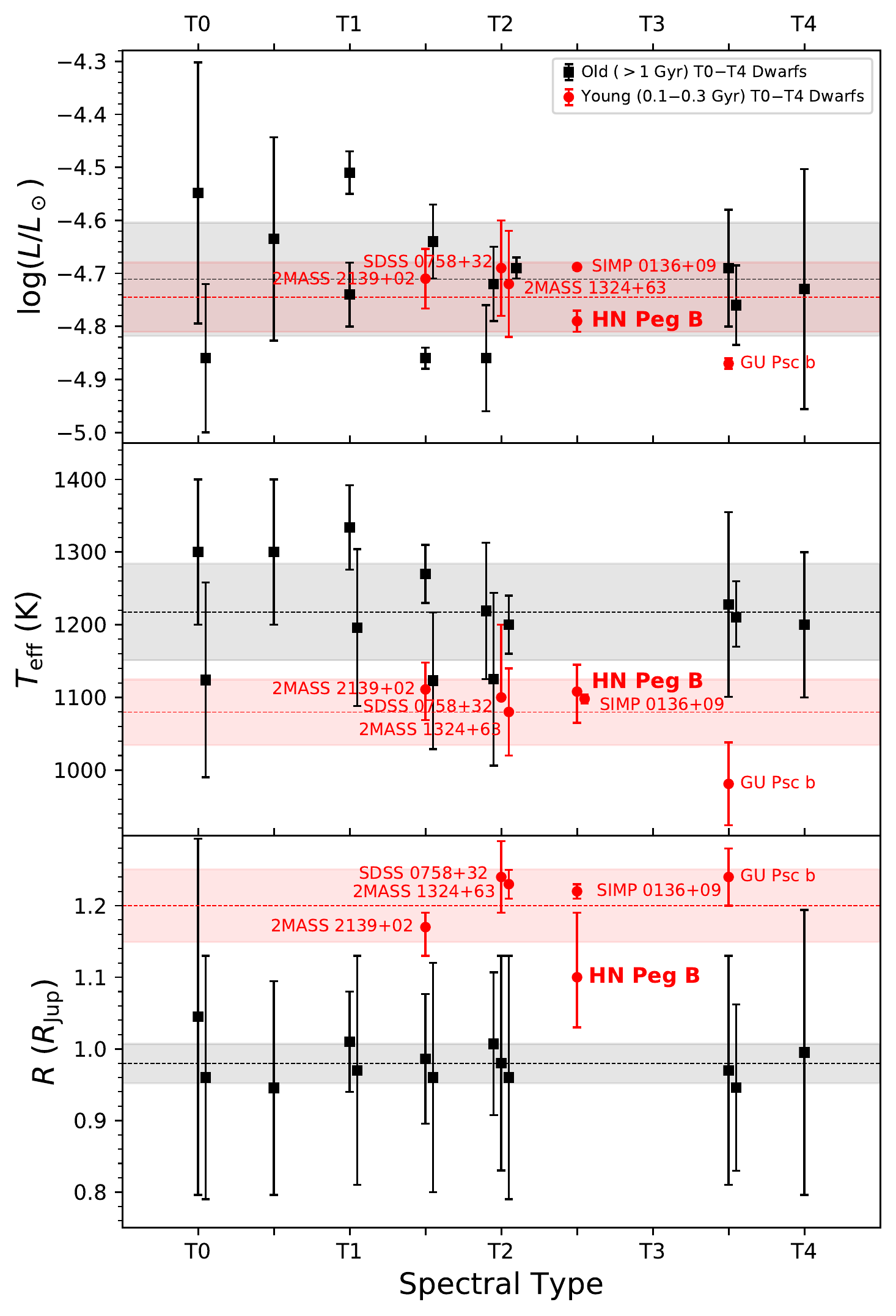}
	\caption{Fundamental parameters $\log L$ (top panel), $T_{\rm eff}$ (middle panel), and $R$ (bottom panel) as a function of spectral type for T0--T4 young ($\approx$0.1--0.3~Gyr; red points) and old ($>$1~Gyr; black squares) dwarfs. The young dwarfs are from \citet[][SDSS J075840.32+324723.3; 40--400 Myr]{Stephens_etal2009}, \citet[][GU Psc b; 50--120~Myr]{Filippazzo_etal2015}, \citet[][SIMP~0136+09; $\sim200\pm50$~Myr]{Gagne_etal2017}, \citet[][2MASS J13243553+6358281; $149_{-19}^{+51}$~Myr]{Gagne_etal2018}, \citet[][2MASS J21392676+0220226; $\sim200\pm50$ Myr]{Zhang_etal2021}, and this work (HN~Peg~B; $300_{-170}^{+280}$ Myr). The sample of old dwarfs was obtained from \citet{Stephens_etal2009}, \citet{Filippazzo_etal2015}, and \citet{Dupuy-Liu2017}. The dashed lines and the shaded regions indicate the average and the standard deviation, respectively, of the parameters of the young (red) and old (black) dwarfs. The averages are: $\log(L/L_{\rm bol})=-4.74\pm0.07$, $T_{\rm eff}=1080\pm45$ K and $R=1.20\pm0.05\ R_{\rm Jup}$ for the young dwarfs and $\log (L/L_{\rm bol})=-4.71\pm0.11$, $T_{\rm eff}=1218\pm66$ K and $R=0.98\pm0.03\ R_{\rm Jup}$ for the old dwarfs. Horizontal offsets of 0.05 spectral type units were applied to dwarfs with the same spectral type to avoid overlap. From SM08 evolutionary models, an object with the average $T_{\rm eff}$, $R$, and $\log L$ parameters for young dwarfs is a $\sim15\ M_{\rm Jup}$ brown dwarf with an age of $\sim$0.2~Gyr, while an object with the mean parameters for old dwarfs is a $\sim45\ M_{\rm Jup}$ brown dwarf with an age of $\sim$2~Gyr.}
	\label{fig:early_T_dwarfs_parameters}
\end{figure*}

Our bolometric luminosity determination ($\log(L_{\rm bol}/L_\odot)=-4.79\pm0.02$) is consistent with the average luminosity ($\log(L_{\rm bol}/L_\odot)=-4.71$, r.m.s.\ scatter of 0.11 dex) of old field T0--T4 dwarfs \citep{Stephens_etal2009,Filippazzo_etal2015,Dupuy-Liu2017}, albeit sitting at the faint end of the range (Figure \ref{fig:early_T_dwarfs_parameters}; top panel). \citet{Luhman_etal2007} found that HN~Peg~B is under-luminous compared to older T2--T3 dwarfs.  However, improved bolometric luminosity estimates with new accurate parallaxes and \textit{WISE} photometry \citep[e.g.,][]{Filippazzo_etal2015,Faherty_etal2016,Best_etal2020}, and discoveries and astrometry of previously unresolved close binaries \citep[e.g.,][]{Dupuy-Liu2012,Dupuy-Liu2017}, have now revised the luminosities of L/T transition dwarfs downwards.  Thus, the luminosity of HN~Peg~B is consistent both with those of old early-T dwarfs in the field and with those of similarly young early-T dwarfs.

Our temperature estimate ($1108_{-43}^{+37}$~K) for HN~Peg~B is 110~K cooler than the average temperature (1218 K with $\sigma=66$ K) for old field T0--T4 dwarfs \citep{Stephens_etal2009,Filippazzo_etal2015,Dupuy-Liu2017}, but consistent with that of other young early-T dwarfs (Figure \ref{fig:early_T_dwarfs_parameters}, middle panel). On average, the young ($\approx$0.1--0.3~Gyr) early-T dwarfs are $140\pm80$~K cooler than their old ($\gtrsim$1~Gyr) field counterparts. This is not necessarily a surprising result, as spectral type is a single empirical measure that subsumes the apparent outcome of complex atmospheric chemistry and dynamics. Indeed, lower effective temperatures are a well-known characteristic of young L/T-transition dwarfs \citep{Metchev-Hillenbrand2006,Luhman_etal2007,Gagne_etal2017} or similarly-aged directly imaged giant planets \citep[e.g.,][]{Marley_etal2012}. 

Correspondingly, the bottom panel of Figure \ref{fig:early_T_dwarfs_parameters} shows that the radius of HN~Peg~B is $\approx$10\% larger than the average radius of old T0--T4 dwarfs \citep{Stephens_etal2009,Filippazzo_etal2015,Dupuy-Liu2017}, and that $\approx$0.1--0.3~Gyr-old T0--T4 dwarfs (including HN~Peg~B) on average have $\approx$20\% larger radii than $>$1~Gyr-old field dwarfs with similar spectral types.

\subsection{Comparison of Findings from Photospheric and Evolutionary Models}
Our photospheric model fits to the full HN~Peg~B spectrophotometric data set (Section~\ref{sec:all_model_comparisons}; Tables~\ref{tab:best-fitting_model_spectra} and \ref{tab:best_model_param}) resulted in effective temperature and surface gravity estimates that are independent of the ones produced by evolutionary models (Section~\ref{sec:phy_par}). Both SM08 and BT-Settl photospheres are consistent with the temperatures from evolutionary models (Table \ref{tab:HN_PegB_parameters}) and with our semi-empirical estimate $T_{\rm eff}=1108_{-43}^{+37}$~K. The SM08 photospheres are also in agreement with the $\log g=4.66_{-0.25}^{+0.20}$ from the evolutionary models.  However, the BT-Settl photosphere fits to the full SED favor a lower surface gravity for HN~Peg~B.  

The semi-empirical effective temperatures and the evolutionary model-based surface gravities are much more precise than the ones from the photospheric fits. This was as also noted in the derivation of the physical properties of the T8 dwarf Gliese 570D by \citet{Geballe_etal2001}. In our case, this is the direct result of our broad spectrophotometric coverage, which includes 98\% of the emitted luminosity by HN~Peg~B, and the very precise \textit{Gaia} distance measurement to HN~Peg~A. While the effective temperature still depends on an accurate estimate of the radius through the geometric dilution factor $\alpha=R/d$ (Equation~\ref{eq:alpha_chi2_r}), and the radius itself relies on a model SED, $\alpha$ is not strongly sensitive to the details of the model SED. Conversely, photospheric model fits to spectra aim to reproduce the atmospheric micro-physics, often at different atmospheric temperatures and pressures simultaneously, and so offer weaker leverage on the macroscopic parameters.  Therefore, we adopt the semi-empirical ($T_{\rm eff}$) or evolutionary model-based ($\log g$) determinations as both more precise and more accurate.

Overall, the BT-Settl photospheric models seem less diagnostic of surface gravity in this moderately young T2.5 dwarf.  Notably, the BT-Settl models rely on one less free parameter (no $f_{\rm sed}$) than the SM08 models.  As we already noted in Section~\ref{sec:all_model_comparisons}, both sets of models face challenges in representing the degree of CH$_4$/CO chemical disequilibrium in HN~Peg~B.  The extra $f_{\rm sed}$ parameter in the SM08 atmospheric models offers a lever to disentangle the effects of dustiness and surface gravity, which is important for young objects. Extra atmospheric dust ($f_{\rm sed}<4$), beyond what may be expected at a moderately low surface gravity, may veil some of the deeper-seated CO. Without an independent cloud parameter that can increase veiling of the spectrum to account for the observed CH4/CO band strengths, the fits with the BT-Settl models are driven to a lower the surface gravity, below what is expected from the evolutionary models. This is the likely reason why BT-Settl photospheric fits to the full SED result in $\sim$1~dex lower surface gravities than the evolutionary models.  Conversely, the additional dust content introduced by $f_{\rm sed}<4$ in the SM08 photospheric models obscures the deeper CO-rich atmospheric layers, leaving mostly the lower-pressure layers visible where, under the assumption of chemical equilibrium, the more volatile CH$_4$ molecule would be more abundant.

\subsection{Mass and Age Estimates}
The evolutionary model-based age of HN~Peg~B obtained in this study ($300_{-170}^{+280}$~Myr) is in agreement with that for the host star HN~Peg~A (\citealt[][$300\pm200$ Myr]{Luhman_etal2007}; \citealt[][$240\pm30$ Myr]{Barnes2007}; \citealt[][200 Myr]{Zuckerman-Song2009}). It is also consistent with the finding of \citet{Leggett_etal2008}, who used model photosphere fits to conclude that the age of HN~Peg~B is toward the older end of the 100--500~Myr range.  \citet{Zhou_etal2018} deduced an age of $\gtrsim$500~Myr, although that is based on the previously undiagnosed sensitivity of the \ion{K}{1} lines to gravity.  Using comprehensive SED spectrophotometry and an accurate distance, we have shown that the evolutionary model-based determinations of the fundamental parameters (including age) are more accurate.  We have also demonstrated that the \ion{K}{1} lines are not usable as a surface gravity indicator for early T dwarfs (Section~\ref{sec:results:ki}).

The corresponding mass of HN~Peg~B is $0.020_{-0.007}^{+0.007}\ M_\odot$ ($21_{-7}^{+7}\ M_{\rm Jup}$). This mass estimate is in full agreement with the original $0.021\pm0.009\ M_\odot$ estimate by \citet{Luhman_etal2007} and the 0.012--0.028~$M_\odot$ constraint by \citet{Leggett_etal2008}.

\section{CONCLUSIONS}
\label{sec:conclusions}
We have presented $Spitzer$ mid-IR photometry and spectra of HN~Peg~B which, together with previously published data, comprise the most complete SED for a young L/T transition dwarf (Figure \ref{fig:SED}). The presented spectrophotometry contains about 98\% of the bolometric luminosity of HN~Peg~B. Using the \textit{Gaia} parallax  for the primary HN~Peg~A, we have obtained accurate and precise estimates for the fundamental parameters of HN~Peg~B (Table \ref{tab:HN_PegB_parameters}).  We confirm independently of its host's stellar age (200--300~Myr) that HN~Peg~B is moderately young ($300_{-170}^{+280}$~Myr).

By analysing various aspects of the individual spectra and the full SED, we have arrived at several principal conclusions.

\begin{enumerate}
\item A comparison of $R$ $\approx$2300 $J$-band Keck/NIRSPEC spectra of T2--T3 dwarfs over a range of ages (and moderate to high gravities) shows that the 1.25~$\mu$m \ion{K}{1} doublet is insensitive to surface gravity. This is distinct from the established strengthening of the absorption of alkali lines with increasing gravity in M and L dwarfs or from the increasing absorption strengths with decreasing gravity in late-T dwarfs.

\item Atmospheric models with condensates or clouds (BT-Settl and SM08) better reproduce the SED of the T2.5 dwarf HN~Peg~B than more recent condensate-free models (Sonora 2018 and ATMO 2020). However, condensate-free models by \citet{Tremblin_etal2019} that incorporate a more general diabatic treatment of convection are able to reproduce the low-resolution near-IR spectrum of HN~Peg~B comparably well.

\item As an early-T dwarf with a moderately low ($\log g=4.7$) surface gravity, HN~Peg~B poses challenges to the current suites of photospheric models in accurately reproducing the observed CH$_4$ and CO absorption strengths.  The effect is most pronounced over the 3--5~$\mu$m region, which spans the strongest CH$_4$ and CO features, but also at the 1.6~$\mu$m, 2.2~$\mu$m, and 7.5~$\mu$m CH$_4$ band heads.  BT-Settl photospheric models (with non-equilibrium chemistry) produce more reliable fits than SM08 models (in chemical equilibrium), but significantly under-estimate the surface gravity in the process. 

\item HN~Peg~B and other young T0--T4 dwarfs are $140\pm80$~K cooler and have $\approx$20\% larger radii compared to old field dwarfs with similar spectral types, but have similar bolometric luminosities. The cooler temperatures at younger ages are in agreement with results for the young L7.5 dwarf HD~203030B \citep{Metchev-Hillenbrand2006,Miles-Paez_etal2017} and for directly imaged young extrasolar giant planets with similar late-L-type photospheres (e.g., HR~8799~cde; \citealt{Marois_etal2008,Marois_etal2010} and YSES 2b; \citealt{Bohn_etal2021}). Hence, the appearance and strengthening of methane at the L/T transition proceeds at cooler temperatures in lower-gravity objects, in concordance with cloudy models \citep{Marley_etal2012}.
\end{enumerate}

{\bf Acknowledgments.} 
The work presented in this paper was supported in part by NASA JPL under awards 1273192 
and 1369094.
Some of the data presented herein were obtained at the W.M.\ Keck Observatory, which is operated as a scientific partnership among the California Institute of Technology, the University of California and the National Aeronautics and Space Administration.  The Observatory was made possible by the generous financial support of the W.M.\ Keck Foundation.

This research has made use of the SVO Filter Profile Service (http://svo2.cab.inta-csic.es/theory/fps/) supported from the Spanish MINECO through grant AYA2017-84089.

\facility{Spitzer, HST, Keck:II, Gemini North, IRTF}
 
\appendix
\section{Comparisons of the HN~Peg B SED to Recent Atmospheric Cloud-free Models}
\label{sec:cloudless_model_fits}
In using the \citet{Saumon-Marley2008} and BT-Settl \citep{Allard_etal2012} atmospheric models to fit the spectra of HN~Peg~B (Section~\ref{sec:model_fits}) we have not taken advantage of models with the most up-to-date molecular line lists and opacities.  The more recent Sonora 2018 \citep{Marley_etal2018} and ATMO 2020 \citep{Phillips_etal2020} models incorporate such improvements, and we compare the SED of HN~Peg~B to these models here.  As we detail below, these more recent models do not match the data as well, most likely because they do not include condensate opacities.

The Sonora 2018 models assume chemical equilibrium. The ATMO 2020 models include three different grids of synthetic spectra, depending on the how non-equilibrium chemistry is treated: full chemical equilibrium (ATMO 2020 CEQ), weak vertical mixing and disequilibrium (ATMO 2020 NEQ weak), and strong vertical mixing and disequilibrium (ATMO 2020 NEQ strong).  Figure~1 in \citet{Phillips_etal2020} shows the vertical mixing relationships with $\log g$. 

We followed the same steps as described for the comparison to SM08 and BT-Settl models in Section~\ref{sec:model_fits}.  We convolved the model photospheres to the resolution of the data and found the best-fitting model photospheres by minimizing $\chi^2_r$ (Equation \ref{eq:chi2_r}).  The grids of parameters for our fits were: $500\le T_{\rm eff}/{\rm{K}}\le2000$ in steps of 100~K and $3.0 \le \log g \le 5.5$ in steps of 0.25~dex for Sonora 2018, and $500\le T_{\rm eff}/{\rm{K}}\le1800$ in steps of 100~K and $2.5 \le \log g \le 5.5$ in steps of 0.5~dex for ATMO 2020. For both model families we used solar metallicity.  Rather than re-doing the detailed spectrum-by-spectrum analysis of Section~\ref{sec:model_fits}, we compared the models only to the full spectrophotometric data set.  As we found in Section~\ref{sec:all_model_comparisons}, this offered the most self-consistent results for the effective temperature and surface gravity of HN~Peg~B.  However, we did exclude the highest-resolution data set: at $J$ band from NIRSPEC. As the $J$ band is among the wavelength regions most strongly affected by clouds, we found that the large number of data points in the NIRSPEC spectrum dominated the fits and drove other wavelength regions further away from fitting well.

\begin{figure*}
	\centering
	\begin{tabular}{cccc}
		\subfloat{\includegraphics[width=.50\linewidth]{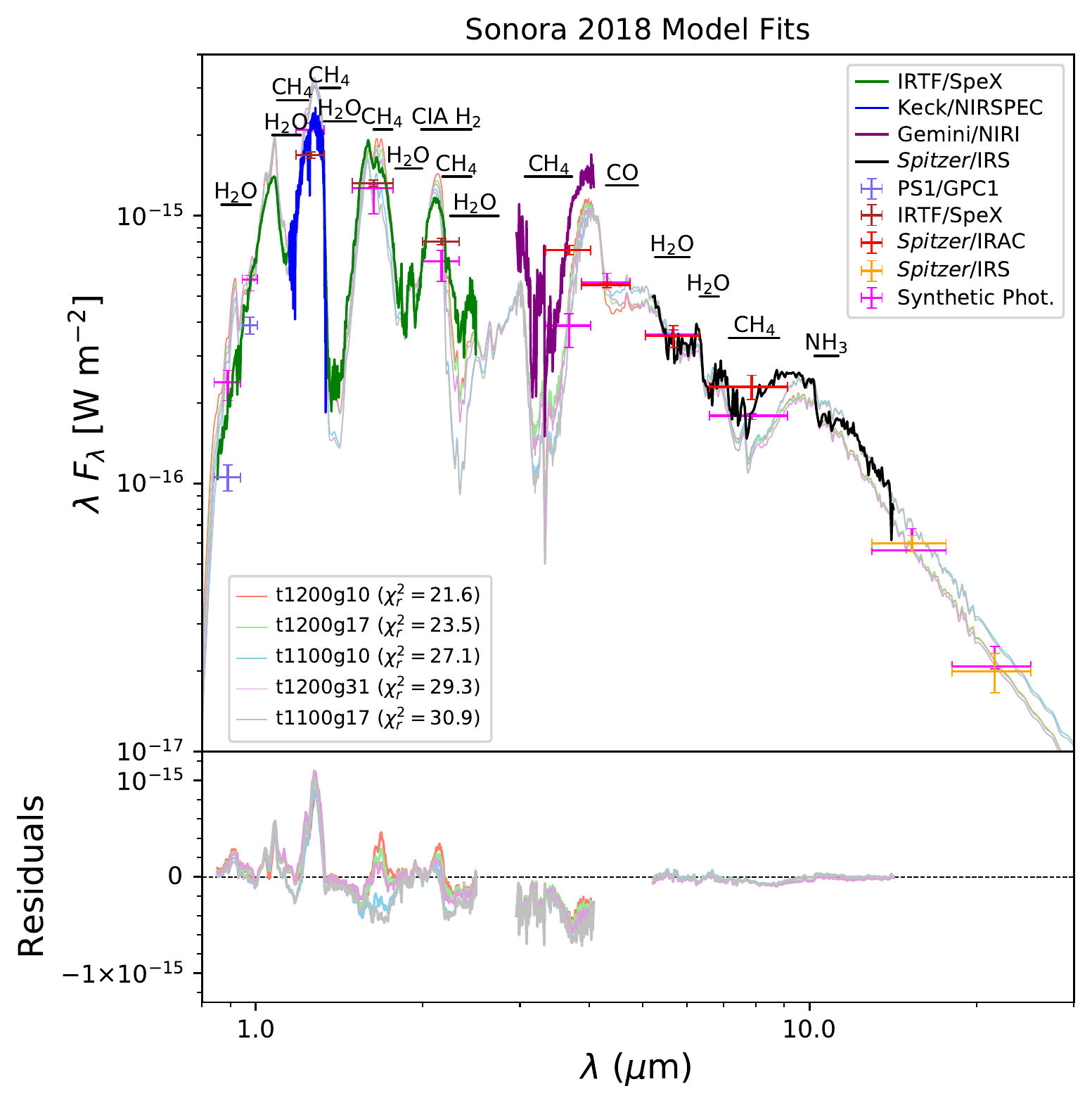}} &
		\subfloat{\includegraphics[width=.50\linewidth]{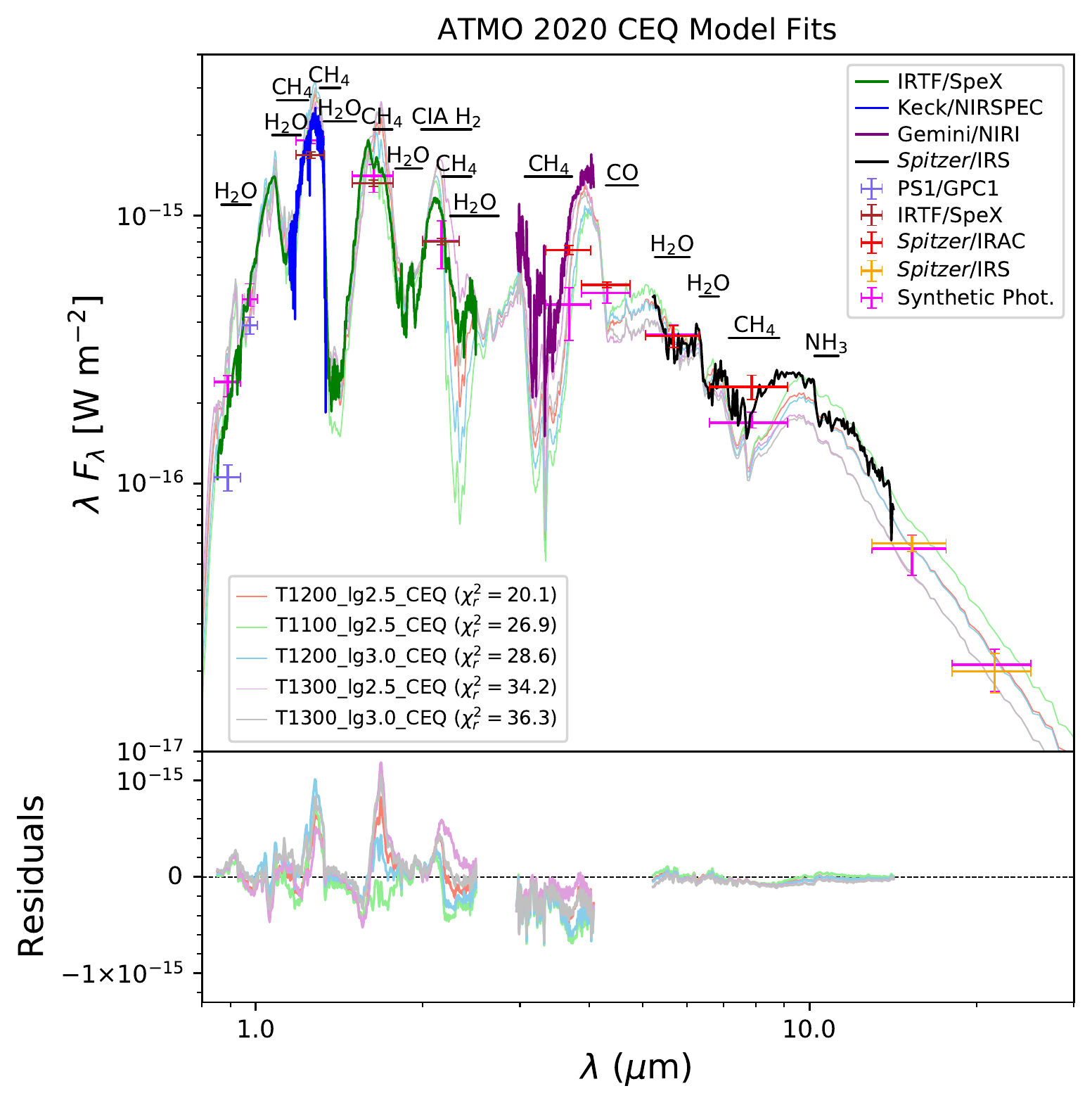}} \\
		\subfloat{\includegraphics[width=.50\linewidth]{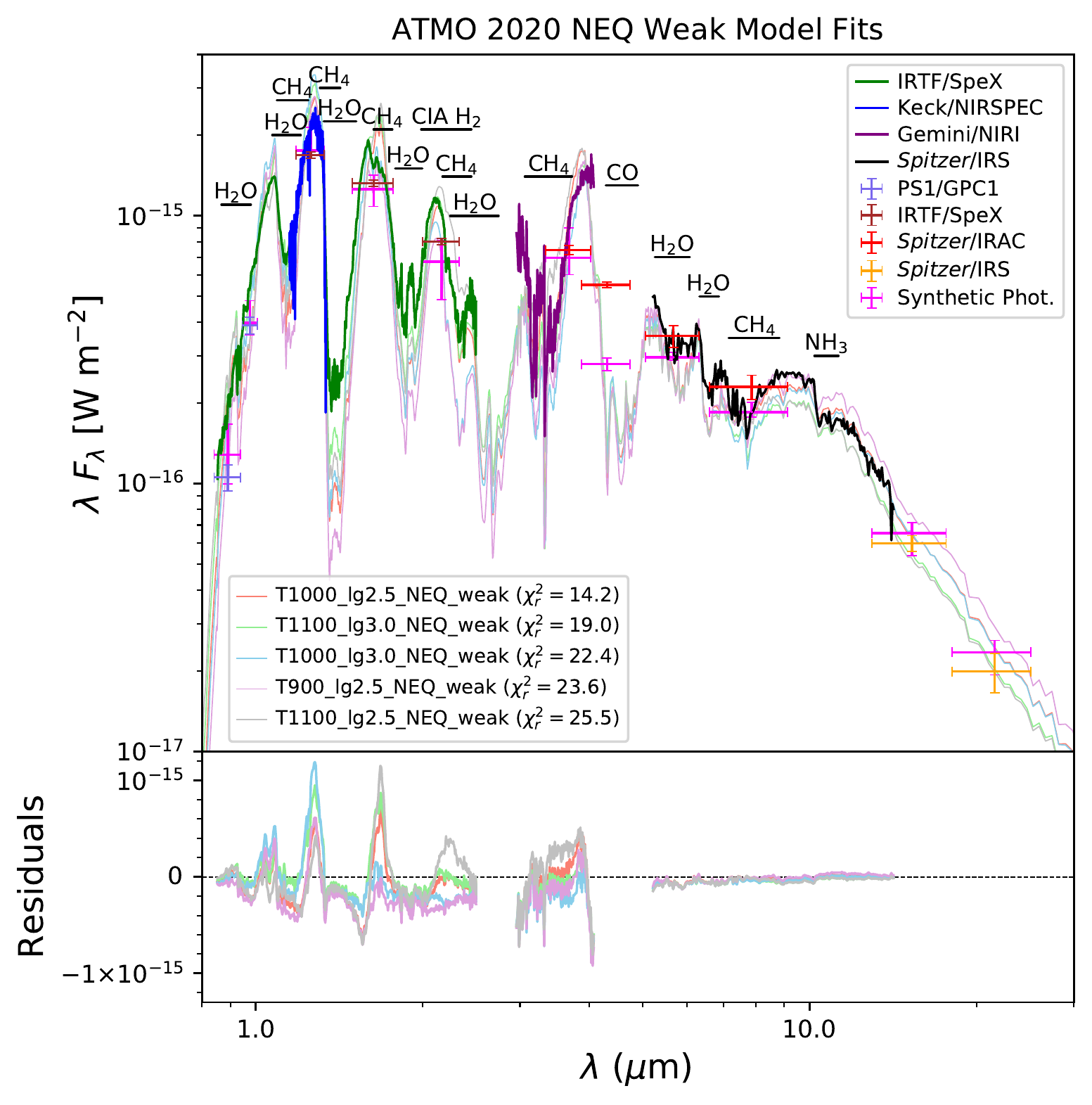}} &
		\subfloat{\includegraphics[width=.50\linewidth]{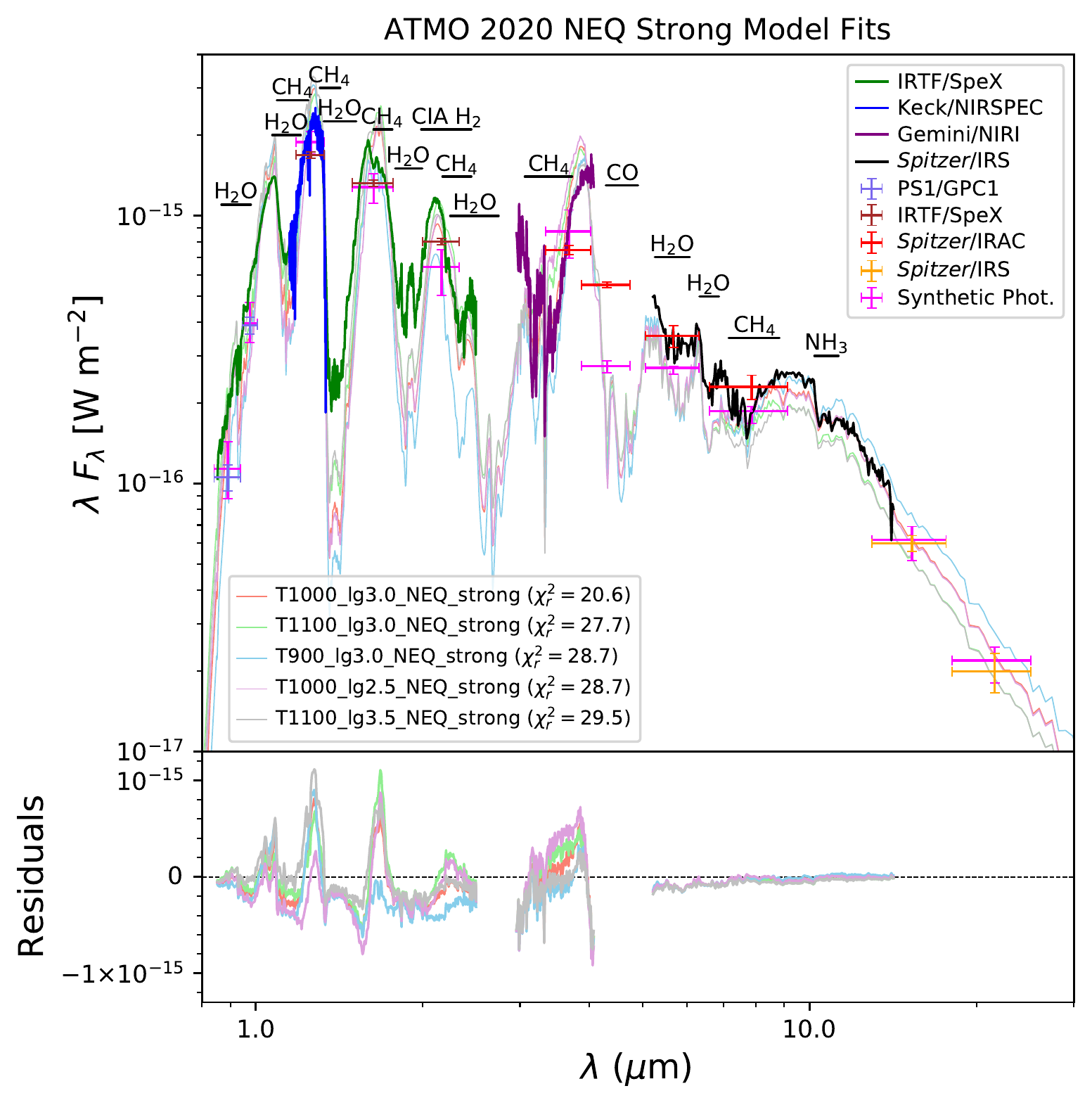}}
	\end{tabular}
	\caption{SED of HN~Peg~B assembled from the spectrophotometry listed in the top-right legends. The best-fitting Sonora 2018 (upper left panel), ATMO 2020 CEQ (upper right panel), ATMO 2020 NEQ weak (lower left panel), and ATMO 2020 NEQ strong (lower right panel) atmospheres to the spectrophotometry (excluding the NIRSPEC $J$-band spectrum) are shown by the curves indicated in the bottom legends ($T_{\rm eff}$ is in K for both model families, and $g$ in m~s$^{-2}$ for Sonora 2018 models and $\log g$ for ATMO 2020 models). The horizontal error bars correspond to the widths of the photometric pass bands (see Table \ref{tab:HN_PegB_phot}). The main spectral features are indicated. The residuals between the best model fits and the data are shown in the bottom panels of each graph.}
	\label{fig:SED_test}
\end{figure*}

We show the five best-fitting Sonora 2018 and ATMO 2020 atmospheres in Figure~\ref{fig:SED_test}. With $\chi^2_r$ values that are 2--3 times larger, these condensate-free models do not perform as well as the cloudy SM08 and BT-Settl models (cf.,\ Table \ref{tab:best_model_param}). 

The best-fitting photospheres with equilibrium chemistry (Sonora 2018 and ATMO 2020 CEQ) over-predict the strength of the methane absorption at both 3--4~$\mu$m and 7--9~$\mu$m. We remarked a similar challenge for the (cloudy) SM08 models that is also likely attributable to the assumption of chemical equilibrium (Section \ref{sec:all_model_comparisons}). Notably however, both Sonora 2018 and ATMO 2020 CEQ models reproduce the strength of the CO absorption at 4.6~$\mu$m well, unlike SM08 and BT-Settl (Section~\ref{sec:all_model_comparisons}).

The best-fitting non-equilibrium chemistry ATMO 2020 models better reproduce the CH$_4$ absorption at 3--4~$\mu$m and 7--9~$\mu$m, but show significantly stronger atmospheric absorption in all H$_2$O bands (Figure~\ref{fig:SED_test}). The water absorption is better fit by the BT-Settl model atmospheres (Figure~\ref{fig:SED}).

The effective temperatures (1100--1200~K) of the best-fit equilibrium chemistry cloudless Sonora 2018 models are similar to the previous generation cloudy SM08 models (1000--1200~K), although the newer models point to a lower surface gravity: $3.0\leq\log g\leq 3.5$ for Sonora 2018 vs.\ $4.0\leq\log g\leq 5.0$ for SM08.  As we found in our spectrophotometrically calibrated estimate of the fundamental parameters of HN~Peg~B in Section~\ref{sec:phy_par}, only the higher surface gravity is consistent with the bolometric estimate ($\log g=4.66_{-0.25}^{+0.20}$; Table~\ref{tab:HN_PegB_parameters}). The best-fit ATMO 2020 models produce effective temperatures (1100--1300~K for CEQ; 900--1100~K for NEQ weak and NEQ strong) that are also consistent with those from BT-Settl (1000--1150~K), although also with lower surface gravities: $2.5\leq\log g\leq 3.5$ for ATMO 2020 vs.\ $3.5\leq\log g\leq 4.0$ for BT-Settl.  All are lower than the $\log g$ estimate from evolutionary models.

Overall, we find that the more recent condensate-free Sonora 2018 and ATMO 2020 models to not reproduce the SED of the partly cloudy HN~Peg~B T2.5 dwarf, despite improvements in the molecular line opacities since the previous generation of SM08 and BT-Settl models.

\bibliography{mybib_Suarez}

\end{document}